# A PERMUTATION APPROACH FOR RANKING OF MULTIVARIATE POPULATIONS

Livio Corain and Luigi Salmaso

# Introduction and motivation

The necessity and usefulness of defining an appropriate ranking of several populations of interests, i.e. processes, products, services, teaching courses, degree programs, and so on is very common within many areas of applied research such as Chemistry, Material Sciences, Engineering, Business, Education, etc. The idea of ranking in fact occurs more or less explicitly all times when in a study the goal is to determine an ordering between several input conditions/treatments with respect to one or more outputs of interest for each at which there is a natural preferable direction.

This happens very often in the context of management and engineering problems where the populations can be products, services, processes, etc. and the inputs are for example the managerial practices or the technological devices which are put in relation with some suitable outputs such as any performance measure. Specific managerial and engineering subjects in which these situations occur are so frequent, let us think on the typical issues of operations management, quality control, and marketing, so that on the sidelines of the problem of ranking a lot of different theories and diverse methods have been developed in the literature. At the same time, the ranking problem is a typical interdisciplinary subject, just think for example on the development process of a new product where managerial practices, engineering issues and statistical techniques are jointly involved in order to achieve high quality and potentially successful products.

Many times the populations of interest are multivariate in nature, meaning that many aspects of that populations can be simultaneously observed on the same unit/subject. For example, in many technological experiments the treatments under evaluation provide an output of tens of even hundreds univariate responses (e.g. think on the myriad of automated measurements that are performed on a silicon wafer during the manufacturing process by microelectronics industry).

Similarly, but in a completely different context, a survey-based observational research provides for each unit/subject (respondent) a long list of answers, as in the case of evaluating and monitoring of customer satisfaction.

From a statistical point of view, when the response variable of interest is multivariate in nature, the inferential problem may become quite difficult to cope with, due to the large dimensionality of the parametric space. Moreover, when the goal is that of comparing several multivariate populations, a further element of difficulty is related to the nature of the response variable. If we consider a continuous response, provided that the underlying distributional and sampling assumptions are met and the degree of freedom are large enough, then inference on populations can be performed using classical methods (e.g. such as Hotelling $T^2$). But when the response variables are ordered categorical the difficulties of the traditional methods based on contingency tables may become insurmountable. Nonparametric inference based on the NPC - NonParametric Combination of several dependent permutation test statistics (see Pesarin and Salmaso, 2010), as we shall see, allows us to overcome these limitations, without the necessity of referring to assume any specified random distribution. The main advantages of using the permutation and combination approach to classify and ranking several multivariate populations is that it is the only one testing method which allow us to derive multivariate directional *p*-values that can be calculated also when the number of response variables are much more larger than the number of replications (so-called *finite-sample consistency* of combined permutation tests). It is worth noting that in this situation, which can be common in many real applications, all traditional parametric and nonparametric testing procedures are not appropriate at all.

After presenting a couple of guideline examples to help the reader to understand in practice the right meaning of the problem of ranking of multivariate populations, this section is devoted to present the formalization and the general solution of the multivariate ranking problem. As will be shown, actually we intend the ranking problem as a non-standard data-driven classification problem, which can be viewed similar to a sort of special case of post-hoc multivariate multiple comparison procedure. In this view the classification procedure is an empirical process that uses pseudo-inferential tools, in particular pseudo-test statistics, with the function of distance indicators and signals useful to rank the populations.

Since the problem of ranking is still addressed in the literature with respect to many other different points of view, in this section we briefly review the basic procedures proposed in the literature, classifying them within the main reference field where they have been developed, that is statistics and operations research. Finally, since our proposed method has little relevance to the usual approaches proposed in the literature around the concept of ranking we close illustrating which are the specificities and advantages of the permutation approach for multivariate ranking problems.

## 1.1  Some guideline examples

Assume that there are several (more than two) populations of interests to be compared one each other in order to establish if they are all equal, i.e. actually if there is just one single population, or if they are different, and in this case we want to classify that populations from the 'best' to the 'worst' according to a given pre-specified criterion. From a general point of view and depending on the specific real context, both experimental and observational studies can be taken into account so that several samples are drawn from populations in order to classify and ranking that populations, where with the term "ranking" we mean a meaningful criterion which allows us to rank populations from the 'best' to the 'worst'. Also assume that the response we observe on each unit/subject is multivariate in nature, where each univariate component can be either continuous or discrete or binary or ordered categorical (we even admit the mixed situation). Finally, we assume that for each univariate component an unique criterion is defined such as it is possible to establish a natural preferable direction (as for instance "the greater the better" or "the smaller the better" or "the closer to the target the better").

### 1.1.1 An experimental-type example

In the first example we consider a real experimental case study with very few replications and where the response is continuous in all univariate components and where the criterion "the higher the better" holds for all components.

In the field of new product development, when developing new detergents the laundry industry refers to the so-called primary detergency, i.e. an experiments devoted to the assessment of benefits of a detergent in removing several types of stains from a piece of previously soiled fabric (Bonnini et al., 2009). When performing a primary detergency experiment, given that the benefits are simultaneously evaluated on several different types of stains (usually ranging from 10 to 30), the response variable can actually be considered as multivariate in nature (Arboretti et al., 2008; Corain and Salmaso, 2007). Figure 1 represents a graphical three-dimensional representation of the observed stain removal performance (so-called reflectance, ranging from 0 to 100) of three populations (products) where for the sake of simplicity we considered just only the first three stains of a primary detergency experiment. The full coloured dots represent the hypothetical true unknown/unobservable product performances (true means) and through this experiment, the four replications are represented by empty circles, we wish to establish if the products have the same performance or if they are different and in this case we want to classify them from the 'best' to the 'worst'. From the simple descriptive inspection of observed data, we expect that the pseudo inferential analysis will suggest that the products are different. The term *pseudo-inference* is used in place of inference because we admit controlling the inferential errors. Indeed, as we will see later on, we use to mime testing procedure just to obtain suitable statistics monotonically related with multivariate distances which would be the true unknown elements on which, if available, the classification process would be based.

Since the response has a natural direction, i.e. the higher the reflectance the better is the product, we also expect that the best product will likely be P1, which looks better than products P2 and P3, which in turn are likely not different and both at the second ranking position. Note that the underlying true ranking can be inferred from the solid lines representing the distances of the true means with the point of absolute theoretical maximum (top right). In fact, P1 is the product much closer to the maximal point.

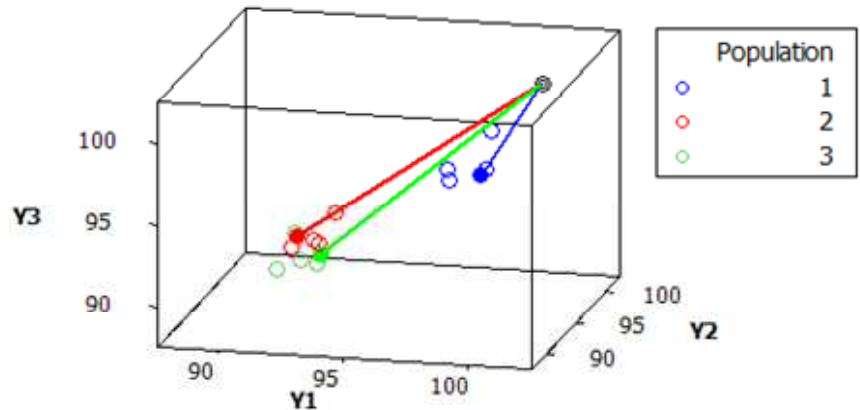

*Figure 1. Graphical three-dimensional representation of a primary detergency experiment.*

It is worth noting that since the populations are multivariate in nature the pseudo inference and the classification process into a global ranking should be properly taken into account for the multivariate distribution of the response variable.

### 1.1.2  An observational-type example

A further element of difficulty of the traditional multivariate inferential methods is related to the nature of the response variable. If we consider a continuous response as in the case of the primary detergency experiment, provided that all assumptions were reasonable met then we could properly use classical likelihood based inferential methods, but when some assumptions were uncertainly met or even when the response variables are ordered categorical the difficulties of the traditional methods based on contingency tables may become insurmountable. Inference based on the NPC method however, allows us to relax most of the stringent assumptions of traditional parametric and nonparametric methods providing a flexible solution which from an inferential view-point is also exact for whatever sample size also in case of ordered categorical responses.

6In order to illustrate an example of ranking of multivariate populations in case of ordered categorical response variable, where observed data are not rigorously obtained by a random sampling procedure, let us refer to an observational survey-based investigation in the field of indoor environmental quality evaluations where a sample of 75 pupils from a primary school have been enrolled (25 subjects for each classroom under investigation). The goal of the study is to rank the three classrooms where the pupils are taught in terms of subjective well-being of indoor environmental quality, related to microclimatic conditions and other building-related factors. For the sake of simplicity let us consider just only two dimensions of the perceived environmental quality, for example an overall score on the thermal and the acoustic comfort, using a Likert scale 1-5. In Table 1 the observed frequencies in the three classrooms are reported.

*Table 1. Frequency table of the perceived thermal and acoustic comfort in the three classrooms.*

| Thermal c.<br>Acoustic c. | School 1 ||||| School 2 ||||| School 3 |||||
|---|---|---|---|---|---|---|---|---|---|---|---|---|---|---|---|
| | 1 | 2 | 3 | 4 | 5 | 1 | 2 | 3 | 4 | 5 | 1 | 2 | 3 | 4 | 5 |
| 1 | - | - | - | - | - | - | - | - | - | - | - | - | - | - | - |
| 2 | - | - | - | - | - | - | - | - | - | - | - | - | - | - | - |
| 3 | - | - | 2 | 3 | 1 | - | - | 3 | 2 | 2 | - | - | 6 | 3 | 1 |
| 4 | - | - | 2 | 6 | 2 | - | - | - | 4 | 2 | - | - | 2 | 5 | 2 |
| 5 | - | - | 1 | 1 | 7 | - | - | 2 | 2 | 8 | - | - | 1 | 2 | 3 |

In order to classify the three classrooms from the best to the worst from the point of view of perceived indoor environmental well-being it is clear that we have to consider the comparisons of the bivariate distributions of the two environmental quality evaluations. Roughly speaking, if the majority of the assessments of a given classroom will stay at the bottom-right cells of the two-way contingencies table more likely that classroom will be classified in a higher rank position. Note that, on the contrary of what happens in the case of numeric/continuous responses, a simple global indicator measuring a sort of distance from the ideal best situation (all frequencies in the bottom-right cell [5,5]) is not here of easily solution.

## 1.2 Formalization of the problem and general solution

To mime the problem as if it were a truly inferential one, assume that data were drawn from each of $C$ multivariate populations (i.e. items/groups/treatments), $C>2$, by means of a pseudo-sampling procedure, so as to make inference on their possible equality and in case of rejection of this hypothesis to classify that populations in order to obtain a relative ranking from the 'best' to the 'worst' according to one pre-specified meaningful criterion. We use the term *relative ranking* because we want to underline that it is not an absolute ranking but an ordering that is only refereed to the *C* populations at hand.

With reference to the so-called one-way MANOVA layout (for more complex designs we refer the reader to the last part 3, section 3.1.1 Type of design), let us formalize the problem within a nonparametric framework: let **Y** be the *p*-dimensional response variable represented as a *p*-vector of the observed data and let us assume, without loss of generality, that large values of each univariate aspect *Y* correspond to a better marginal performance and therefore to a higher ranking position. In other words, we are assuming the criterion "the larger the better". The term "large values" has a clear meaning in case of continuous responses, while in case of binary or ordered categorical responses should be intended in terms of "large proportion" and of "large frequencies of high score categories" respectively. The marginal univariate components of **Y** are not restricted to belong to the same type, in other words we can consider also the situation of mixed variables (some continuous/binary and some other ordered categorical).

We recall that our goal is to classify and ranking *C* multivariate populations with respect to *p* marginal variables where samples of $n_j$ independent replicates, $j=1,...,C$, from each population are available.

Under the hypothesis of distributional equality of the *C* populations, all true global rankings would necessarily be equal to one, hence they would be in a full ex-aequo situation. This situation of equal ranking where all populations belong to just one class may be formally represented in a testing-like framework where the hypotheses of interest are:

$$\begin{cases} H_0 : \mathbf{Y}_1 \stackrel{d}{=} \mathbf{Y}_2 \stackrel{d}{=} ... \stackrel{d}{=} \mathbf{Y}_C \\ H_1 : \exists \mathbf{Y}_j \stackrel{d}{\neq} \mathbf{Y}_h, j,h=1,...,C, j \neq h \end{cases}.$$  (1)

In case of rejection of the global multivariate hypothesis $H_0$, that is when data are evidence of the fact at least one population behaves differently from the others, it is of interest to perform pseudo-inferences on pairwise comparisons between populations, i.e.

$$\begin{cases} H_{0(jh)}: \mathbf{Y}_j \stackrel{d}{=} \mathbf{Y}_h \\ H_{1(jh)}: \mathbf{Y}_j \stackrel{d}{\neq} \mathbf{Y}_h, j,h=1,...,C, j \neq h \end{cases}. \quad (2)$$

Note that a rejection of at least one hypothesis $H_{0(jh)}$ implies that we are not in an equal ranking situation, that is at least one multivariate population has a greater global rank than some others. Finally, to make pseudo-inference on which marginal variable(s) that inequality is mostly due to, it is useful considering the pseudo-inferences on univariate pairwise comparisons between populations, defined as:

$$\begin{cases} H_{0k(jh)}: Y_{jk} \stackrel{d}{=} Y_{hk} \\ H_{1k(jh)}: \left(Y_{jk} \stackrel{d}{<} Y_{hk}\right) \bigcup \left(Y_{jk} \stackrel{d}{>} Y_{hk}\right) \end{cases}, j,h=1,...,C, j \neq h, k=1,..,p \ , \quad (3)$$

because when $Y_{jk} \stackrel{d}{\neq} Y_{hk}$ is true, then one and only one between $Y_{jk} \stackrel{d}{<} Y_{hk}$ and $Y_{jk} \stackrel{d}{>} Y_{hk}$ is true, i.e. they cannot be jointly true.

Looking at the univariate alternative hypothesis $H_{1k(jh)}$, note that we are mostly interested in deciding whether a population is either greater or smaller than another one (not only establishing if they are different). In this connection, we can take account separately of the directional type alternatives, namely those that are suitable for testing both one-sided alternatives (see Pesarin and Salmaso, 2010, p. 163; Bertoluzzo et. all, 2012). In this connection, also expression (2) can be reformulated as

$$\begin{cases} H_{0(jh)}: \mathbf{Y}_j \stackrel{d}{=} \mathbf{Y}_h \\ H_{1(jh)}: \left(\mathbf{Y}_j \stackrel{d}{<} \mathbf{Y}_h\right) \bigcup \left(\mathbf{Y}_j \stackrel{d}{>} \mathbf{Y}_h\right), j,h=1,...,C, j \neq h \end{cases}. \quad (2\text{bis})$$

Let be $p^+_{k(jh)}$ and $p^-_{k(jh)}$ the marginal directional *p*-value-like statistics related to the stochastic inferiority or superiority alternatives $H^+_{1k(jh)}: Y_{jk} \overset{d}{>} Y_{hk}$ and $H^-_{1k(jh)}: Y_{jk} \overset{d}{<} Y_{hk}$, respectively. Since by definition $p^+_{k(jh)} = 1 - p^-_{k(jh)} = p^-_{k(hj)}$, note that all one-sided pseudo-inferential results related to the hypotheses (3) can be represented as follows:

$$P^+ = \begin{bmatrix} - & p^+_{1(12)} & p^+_{1(13)} & \cdots & p^+_{1(1C)} \\ p^+_{1(21)} & - & p^+_{1(23)} & \cdots & p^+_{1(2C)} \\ \cdots & \cdots & - & \cdots & \cdots \\ p^+_{1(C-1,1)} & p^+_{1(C-1,2)} & \cdots & - & p^+_{1(C-1,C)} \\ p^+_{1(C1)} & p^+_{1(C2)} & \cdots & p^+_{1(C,C-1)} & - \end{bmatrix}, \ldots, \begin{bmatrix} - & p^+_{p(12)} & p^+_{p(13)} & \cdots & p^+_{p(1C)} \\ p^+_{p(21)} & - & p^+_{p(23)} & \cdots & p^+_{p(2C)} \\ \cdots & \cdots & - & \cdots & \cdots \\ p^+_{p(C-1,1)} & p^+_{p(C-1,2)} & \cdots & - & p^+_{p(C-1,C)} \\ p^+_{p(C1)} & p^+_{p(C2)} & \cdots & p^+_{p(C,C-1)} & - \end{bmatrix}. \quad (4)$$

Finally, let be $p^+_{\bullet(jh)}$ the directional *p*-value-like statistics calculated via nonparametric combination methodology and related to the multivariate stochastic superiority alternatives $H^+_{1(jh)}: \mathbf{Y}_j \overset{d}{>} \mathbf{Y}_h$ of expression (2bis). All the $C \times (C-1)$ $p^+_{\bullet(jh)}$ can be represented as follows:

$$P^+_\bullet = \begin{bmatrix} - & p^+_{\bullet(12)} & p^+_{\bullet(13)} & \cdots & p^+_{\bullet(1C)} \\ p^+_{\bullet(21)} & - & p^+_{\bullet(23)} & \cdots & p^+_{\bullet(2C)} \\ \cdots & \cdots & - & \cdots & \cdots \\ p^+_{\bullet(C-1,1)} & p^+_{\bullet(C-1,2)} & \cdots & - & p^+_{\bullet(C-1,C)} \\ p^+_{\bullet(C1)} & p^+_{\bullet(C2)} & \cdots & p^+_{\bullet(C,C-1)} & - \end{bmatrix}. \quad (5)$$

Note that *p*-value-like statistics in expression (5) indicate either if there is a possible global multivariate significant dominance between each pairs of populations and in which global direction this dominance can actually exist. It is worth noting that, on the contrary to what happens for the marginal directional *p*-value like statistics, the constraint to sum to one does not hold in this case, i.e. $p^+_{\bullet(jh)} \neq 1 - p^-_{\bullet(jh)}$.

Now, let us consider a last set of *C* multivariate global directional *p*-value-like statistics $p^+_{\bullet(j\bullet)}$, *j*=1,...,*C*, calculated via nonparametric combination methodology and related to the alternatives

$$H_{1(j\bullet)}^{+}: \bigcup_h \mathbf{Y}_j \overset{d}{>} \mathbf{Y}_h, h=1,...,C, j \neq h. \tag{6}$$

These non-standard and unusual *p*-value-like statistics play the role of a sort of *multivariate score* because actually they measure the evidence in favour of the overall stochastic dominance of each populations *against all others*: in fact, the lower $p_{\bullet(j\bullet)}^{+}$, *j*=1,...,*C*, the higher is the evidence of the global dominance of the *j*-th population when pairwise compared with all others. Let us order from the smallest to the largest the set of the $p_{\bullet(j\bullet)}^{+}$, *j*=1,...,*C*: $p_{\bullet[1\bullet]}^{+}$,..., $p_{\bullet[C\bullet]}^{+}$, then we can apply the same ranking [1],....[C] to the matrix $P_{\bullet}^{+}$, defined as the *C*×*C* matrix of the unordered $p_{\bullet(j\bullet)}^{+}$, then we remove the lower- diagonal elements to define a final upper diagonal *p*-value-like statistic matrix $P_{[\bullet]upper}^{+}$:

$$P_{[\bullet]}^{+} = \begin{bmatrix} - & p_{\bullet[12]}^{+} & p_{\bullet[13]}^{+} & \cdots & p_{\bullet[1C]}^{+} \\ p_{\bullet[21]}^{+} & - & p_{\bullet[23]}^{+} & \cdots & p_{\bullet[2C]}^{+} \\ \cdots & \cdots & - & \cdots & \cdots \\ p_{\bullet[C-1,1]}^{+} & p_{\bullet[C-1,2]}^{+} & \cdots & - & p_{\bullet[C-1,C]}^{+} \\ p_{\bullet[C1]}^{+} & p_{\bullet[C2]}^{+} & \cdots & p_{\bullet[C,C-1]}^{+} & - \end{bmatrix} \Rightarrow P_{[\bullet]upper}^{+} = \begin{bmatrix} - & p_{\bullet[12]}^{+} & p_{\bullet[13]}^{+} & \cdots & p_{\bullet[1C]}^{+} \\ - & - & p_{\bullet[23]}^{+} & \cdots & p_{\bullet[2C]}^{+} \\ - & - & - & \cdots & \cdots \\ - & - & - & - & p_{\bullet[C-1,C]}^{+} \\ - & - & - & - & - \end{bmatrix}. \tag{7}$$

Note that expression (7) simply performs a rearrangement of the matrix $P_{\bullet}^{+}$ moving its rows and columns in order to put in the first rows the "best" estimated populations. In practice, this procedure is not nothing but a data-driven selection of *p*-values in (5) with the following rationale: first we pre-order the populations using $p_{\bullet(j\bullet)}^{+}$ which is an indicator of the overall dominance of each population when compared to all other then we summarize the evidence of the relevant *C*×(*C*−1)/2 pairwise stochastic dominances, removing half of the not informative *p*-values.

Once a suitable adjustment for multiplicity has been applied to $P_{[\bullet]upper}^{+}$ and a given pseudo-significance level has been chosen, the classification of the *C* multivariate populations into a final global ranking can be very easy performed but just in only two special cases: when all adjusted *p*-value-like statistics are significant or when they are significant in blocks (i.e. significant *p*-value-like statistics are arranged in blocks, so that there is evidence in favor of "cluster of populations"). In all other situations, i.e. when not all *p*-value-like statistics are significant or any significant blocking arrangement does not exist, a suitable algorithm is needed in order to univocally derive from expression (7) a final global ranking. Details on this algorithm can be found in part 3, paragraph 3.2 "From directional multivariate pairwise comparisons to global ranking".

To summarize the basic ideas behind the formalization of the multivariate ranking problem, note that we start from a multivariate *C*-sample problem to actually define a non-standard data-driven classification and ranking problem, which can be viewed to mime a sort of special case of post-hoc multivariate multiple comparison procedure.

To make more clear the proposed procedure, let us apply it to the two guideline examples presented in the previous sub-section 1.1. Tables 2-5 represent $P^+$, $P_\bullet^+$ and $P^+_{[\bullet]upper}$ i.e. the univariate marginal and the multivariate unordered and ordered directional permutation *p*-value-like statistics, where the last one has been adjusted by multiplicity using the Bonferroni-Holm-Shaffer method.

*Table 2. Marginal directional permutation p-value-like statistics for the first guideline example (three product primary detergency experiment).*

|       |    | Stain 1 |       |       | Stain 2 |       |       | Stain 3 |       |       |
|-------|----|---------|-------|-------|---------|-------|-------|---------|-------|-------|
|       |    | P1      | P2    | P3    | P1      | P2    | P3    | P1      | P2    | P3    |
|       | P1 | -       | .0021 | .0012 | -       | .0018 | .0011 | -       | .0007 | .0045 |
| $P^+ =$ | P2 | .9979   | -     | .6710 | .9982   | -     | .3418 | .9993   | -     | .5519 |
|       | P3 | .9988   | .3290 | -     | .9989   | .6582 | -     | .9955   | .4481 | -     |

*Table 3. Multivariate directional p-value-like statistics for the first guideline example (three product primary detergency experiment).*

|           |    | P1    | P2    | P3    |   | $p^+_{\bullet(j\bullet)}$ |       |           |    | P1 | P3    | P2    |
|-----------|----|-------|-------|-------|---|---------------------------|-------|-----------|----|----|-------|-------|
|           | P1 | -     | .0001 | .0002 |   | P1                        | .0001 |           | P1 | -  | .0002 | .0003 |
| $P^+_\bullet =$ | P2 | .9984 | -     | .8110 |   | P2                        | .8120 | $P^+_{[\bullet]} =$ | P3 | -  | -     | .7510 |
|           | P3 | .9841 | .7510 | -     |   | P3                        | .7942 |           | P2 | -  | -     | -     |

*Table 4. Marginal directional permutation p-value-like statistics for the second guideline example (perceived comfort in the three classrooms).*

|         |    | Thermal comfort |       |       | Acoustic comfort |       |       |
|---------|----|-----------------|-------|-------|------------------|-------|-------|
|         |    | C1              | C2    | C3    | C1               | C2    | C3    |
|         | C1 | -               | .0021 | .0012 | -                | .0018 | .0011 |
| $P^+ =$ | C2 | .9979           | -     | .6710 | .9982            | -     | .3418 |
|         | C3 | .9988           | .3290 | -     | .9989            | .6582 | -     |

*Table 5. Multivariate directional p-value-like statistics for the second guideline example (perceived comfort in the three classrooms).*

|           |    | C1    | C2    | C3    |   | $p^+_{\bullet(j\bullet)}$ |       |           |    | C1 | C3    | C2    |
|-----------|----|-------|-------|-------|---|---------------------------|-------|-----------|----|----|-------|-------|
|           | C1 | -     | .0001 | .0002 |   | C1                        | .0001 |           | C1 | -  | .7510 | .0003 |
| $P^+_\bullet =$ | C2 | .9984 | -     | .8110 |   | C2                        | .8120 | $P^+_{[\bullet]} =$ | C3 | -  | -     | .0002 |
|           | C3 | .9841 | .7510 | -     |   | C3                        | .7942 |           | C2 | -  | -     | -     |

From the adjusted $P^+_{[\bullet]upper}$ it is very easy to classify the three products (first example) and the three classrooms (second example) as follow:

- global ranking of the three detergents: P1=1, P2=2, P3=2;
- global ranking of the three classrooms: Classroom 1=1, Classroom 2=1, Classroom 3=3.

In fact, note that in both cases the significant *p*-value-like statistics (α=5%) are actually arranged in blocks.

## 1.3 Specificities and advantages of the permutation approach for the multivariate ranking problem

A critical revision of the literature on the ranking problem highlights that despite the fact the problem has been extensively treated from many perspectives, none of these seems to be close enough to the procedure we proposed neither from the point of view of the purpose, nor the method used. Even if Multiple Comparison Procedures (MCPs) address the problem of ranking the treatment groups (Westfall et al., 1999), there is generally, no clear indication as to how to deal with the information from pairwise comparisons, especially in the case of a multivariate response variable that is completely ignored. The ranking and selection approach in multiple decision theory, as can be seen in Gupta and Panchapakesan (2002) which contains an extensive discussion on the whole theory, provides some hints, but these are essentially for univariate problems and under assumption of normality. Furthermore, although that book considers a great number of available procedures, it is more focused on theoretical aspects rather than providing practical rules that can be applied directly in real situations. Finally, the vast literature in operations research cannot be of any help for our purposes because even if it comes to establish decision-making algorithms the reference context is never the inference, that of decision making under uncertainty.

As mentioned in Section 1.2 where we presented the formalization of the multivariate ranking problem, it is clear that its general solution requires a key element: an pseudo hypothesis testing procedure for directional multivariate alternatives. To the best of our knowledge, the only method proposed by literature that achieves this goal is the nonparametric combination of dependent permutation tests, the so-called NPC methodology (Pesarin and Salmaso, 2010). The main advantages of using the permutation and combination approach to classify and ranking several multivariate populations is that it allows us to derive multivariate directional $p$-value-like statistics that can be calculated also when the number of response variables are much more larger than the number of replications (so-called *finite-sample consistency* of combined permutation tests). It is worth noting that in this situation, which can be common in many real applications, all traditional parametric and nonparametric testing procedures are not appropriate at all. The NPC approach has a lot of nice feature: it always provides an exact solution for whatever sample size, is very low demanding in terms of assumptions and finally is quite flexible because it allows to jointly handle with all type of response variables, i.e. numeric, binary and ordered categorical even in the presence of any non-informative or informative missing data (missing completely at random or not at random).

From the point of view of applied research, the multivariate ranking problem can be viewed as a tool of quality improvement. In fact, as we mentioned in the first section of this section we remarked that the necessity and usefulness of defining an appropriate ranking of items quite often occurs as a natural conclusion of many researches in the real world of managerial and engineering activities. It is worth noting that in this context the ranking problem takes on a connotation of an effort aimed at obtaining some form of progress or improvement of an organizational aspect or a business practice or a process/product technology. In other words, it can be viewed as a quality improvement task. To support this argument, let us reconsider the two guideline examples: the second one was related to a real case study in the field of indoor environmental quality evaluations where a sample of pupils from a primary school have been enrolled in order to fulfil a questionnaire and at the same time several instrumental measures have been recorded. The goal of the study is to compare and rank three classrooms where pupils attend lessons in terms of subjective and objective well-being of indoor environmental quality, related to microclimatic conditions and building-related factors. Actually the three classrooms differ one each other because of their specific layout, exposure to the sun, position of the windows, etc. The classroom ranking in terms of environmental quality evaluations will be useful in linking the wellness to their specific characteristics to derive in this way useful information for improving the quality of indoor environments. Let us reconsider also the first more technical/technological guideline example in the field of new product development for the laundry industry. When developing new detergents a special kind of washing experiments called primary performance detergency tests are usually performed. These trials are devoted to the assessment of benefits of a detergent in removing a stain from a piece of previously soiled fabric. Also in this context the ranking problem of several detergents under investigation (usually some new products and a presently market product as benchmark) can be viewed as a comparative study where the interest is not only to establish if the products are equal or different but also to find out a suitable ranking (possibly by including some ties) able to evaluate the relative degree of preference of a given products (treatment/condition) with respect to all others. Note that the washing case study arises from a well designed experiment so that the underlying inference is truly effective. On the contrary, in the indoor quality survey we are facing an observational case study so that the related inference are necessarily weaker.

It is worth noting that quite often the items of interest to be ranked are multivariate in nature, meaning that many aspects of the items can be simultaneously observed on the same unit/subject. In the indoor environmental quality study a lot of different subjective evaluations (thermal sensation, acoustic comfort, light intensity, etc.) and objective measures (temperature, humidity, etc.) are jointly recorded. Similarly, when performing a primary detergency experiment, since the benefits are simultaneously evaluated on several different types of stains (grass, coffee, juice, tomato, etc.), the response variable is multivariate in nature.

This consideration leads us to try to define and characterize a new class of multivariate ranking problems. Assuming that a true underling ranking of the items under investigation does exist, using some suitable multivariate sampling information we would like to take a pseudo decision about the possible equality of all items (that is all items are tied) versus a procedure of estimation of a suitable ranking. This goal represents a non standard statistical pseudo inferential problem where both hypothesis testing and classification are jointly involved. At the same time, since we are referring to a multivariate setting this issue represents also a complex problem so that a flexible nonparametric solution is advisable.

In the next section we will suggest a general nonparametric permutation and combination-based theoretical framework (Pesarin and Salmaso, 2010) where new solutions for the ranking problem can be developed.

# Permutation tests and nonparametric combination methodology

In the previous section we presented the theoretical background on ranking of multivariate populations where the main issue of the proposed method was concerned with a suitable set of multivariate pairwise one-sided hypothesis testing procedures. In this section we will show that, thanks to its componentwise nature and to its flexibility to handle with either numeric and ordered categorical data, the combined permutation tests represent a valid solution for the problem at hand. This section aims also at introducing the reader to the theory of univariate and multivariate permutation tests showing the advantage of using such a nonparametric procedures instead of traditional parametric solutions.

## 1.5 Introduction to permutation tests

The importance of the permutation approach in resolving a large number of inferential problems is well-documented in the literature, where the relevant theoretical aspects emerge, as well as the extreme effectiveness and flexibility, from an applicatory point of view (Basso et al., 2009; Edgington and Onghena, 2007; Good, 2010; Pesarin and Salmaso, 2010).

When compared with the more traditional parametric or nonparametric rank-based solutions, the main advantages of using the permutation approach in hypothesis testing problems are that in general permutation tests require fewer and easy to justify assumptions, are exact in nature and offer flexible solutions in dealing with complex problems. In this respect, the permutation-based solution for a complex problem such as the comparison of interventions in group randomized trials (GRT's), is able to maintain a nominal test size thanks to its intrinsic exactness also in case of small sample sizes that usually occur in real applications but are not sufficient to make possible using asymptotic approximations (Braun and Feng, 2001). In the same paper, a simulation study proves that in the case of the usual realistic sample sizes some traditional asymptotic-based procedures for the problem at hand (GEE - generalized estimating equations) have liberal sizes, i.e. they do not maintain the nominal level. Moreover, when considering a suitable model-based testing procedure (PQL - penalized quasi likelihood) even if it is slightly more powerful than the permutation tests when the model of the simulated data exactly corresponds to that assumed, it is outperformed by the permutation tests when there are too few clusters to support asymptotic methods. In summary, permutation tests for GRT's are appropriate and solutions and more general powerful than asymptotic counterparts in that they require fewer distributional assumptions. The use of permutation tests is becoming increasingly popular in biomedical research thanks in part to the effective debate within the community of biostatisticians. In a popular paper of Ludbrook and Dudley (1998), authors argued that, since randomization rather than random sampling is the norm in biomedical research and because group sizes are usually small, exact permutation or randomization tests for differences in location should be preferred over t or F tests. In this connection, when selecting the more appropriate test statistic and in the planning of the size of a study, Weinberg and Lagakos (2000) derive the asymptotic distribution of permutation tests under a general contiguous alternative, and then investigate the implications for test selection and study design for several diverse areas of biomedical applications.

Del Castillo and Colosimo (2011) propose a permutation test for detecting differences in shape for the analysis of experiments where the response is the geometric shape of a manufactured part. They showed that the permutation test provides higher power for 2D circular profiles than the traditional F-based methods used in manufacturing practice, which are based on the circularity form of errors. Authors highlight that the proposed permutation test does not require the error assumptions that are needed in the F test, which may be too restrictive in practice. Still, in the context of the shape analysis, but more from a biological and morphometric point of view, Iaci et al. (2008) propose a permutation-based significance test for a new general index based on Kullback–Leibler information that measures relationships between multiple sets of random vectors.

In the context of statistical analysis for spatial point patterns Ute (2012) proposes a studentized permutation test for the null hypothesis that two (or more) observed point patterns are realizations of the same spatial point process model. The proposed test performs convincingly well in terms of empirical level and appears more powerful than a bootstrap-type competitor proposed in the literature. The superiority of the permutation tests toward bootstrap solutions is proved also by Troendle et al. (2004) when testing the equality of two multivariate distributions for small sample sizes and in the case of high dimension such as when analysing microarray data. When the interest is in detecting genes that are possibly expressed in only a part of the cases or expressed at different levels among the cases (so-called over-expression), van Wieringen et al. (2008) proposed a new permutation-type test based on the mixing proportion in a nonparametric mixture and minimizes a weighted distance function. They proved by a simulation study that this permutation test is indeed more powerful than the two-sample t test and the Cramér–von Mises test.

Sometimes problems of interest are so complex that an asymptotic procedure is too complicated to be developed and the related rate of convergence too difficult to determine and thus it is preferred to propose a permutation tests, which thanks to their simplicity and flexibility can often offer a possible effective solution. For example Cook and Yin (2001) suggest a permutation test as a means of making inference for dimension reduction in discriminant analysis. Hothorn et al. (2006) propose new theoretical framework for permutation tests that opens up the way to a unified and generalized view, emphasizing the flexibility of permutation tests as conditioned testing procedures, where conditioning is with respect to the observed data which are always sufficient statistics in the null hypothesis for any underlying distribution (Pesarin and Salmaso, 2010). Even when some normal theory-type solutions for a given multivariate testing problem such as MANOVA F-tests does exist, the assumption of multivariate normality is often violated in practice, and the impact of such a violation on the validity of tests may be greater when the sample size is smaller (Zeng et al., 2011). Thus, for most sample sizes of practical interest, the relative lack of efficiency of permutation solutions may sometimes be compensated by the lack of approximation of parametric asymptotic counterparts. In addition, assumptions regarding the validity of parametric methods (such as normality and random sampling) are rarely satisfied in practice, so that consequent inferences, when not improper, are necessarily approximated, and their approximations are often difficult to assess.

For any general testing problem, in the null hypothesis ($H_0$), which usually assumes that data come from only one (with respect to groups) unknown population distribution $P$, the whole set of observed data $\mathbf{x}$ is considered to be a random sample, taking values on sample space $\mathcal{X}^n$, where $\mathbf{x}$ is one observation of the $n$-dimensional sampling variable $\mathbf{X}^{(n)}$ and where this random sample does not necessarily have independent and identically distributed (i.i.d.) components, in fact it suffices that data are exchengeable. We note that the observed data set $\mathbf{x}$ is always a set of sufficient statistics in $H_0$ for any underlying distribution.

Given a sample point $\mathbf{x}$, if $\mathbf{x}^* \in \mathcal{X}^n$ is such that the likelihood ratio $f_P^{(n)}(\mathbf{x}) / f_P^{(n)}(\mathbf{x}^*) = \rho(\mathbf{x},\mathbf{x}^*)$ is not dependent on $f_P$ for whatever $P \in \mathcal{P}$, then $\mathbf{x}$ and $\mathbf{x}^*$ are said to *contain essentially the same amount of information with respect to P*, so that they are equivalent for inferential purposes. The set of points that are equivalent to $\mathbf{x}$, with respect to the information contained, is called the orbit associated with $\mathbf{x}$, and is denoted by $\mathcal{X}^n_{/\mathbf{x}}$, so that $\mathcal{X}^n_{/\mathbf{x}} = \bigcup_{u^* \in \prod(u)} \left( \mathbf{X} u_i^*, i=1,...,n; n_1, n_2 \right)$ where $\prod(u)$ is the set of all permutations of $u=(1,...,n)$.

The same conclusion is obtained if $f_P^{(n)}(\mathbf{x})$ is assumed to be invariant with respect to permutations of the arguments of $\mathbf{x}$; i.e., the elements $(x_1,...,x_n)$. This happens when the assumption of independence for observable data is replaced by that of *exchangeability*, $f_P^{(n)}(x_1,...,x_n) = f_P^{(n)}(x_{u^*_1},...,x_{u^*_n})$, where $(u^*_1,...,u^*_n)$ is any permutation of $(1,...,n)$. Note that, in the context of permutation tests, this concept of exchangeability is often referred to as the *exchangeability of the observed data with respect to groups.* Orbits $\mathcal{X}^n_{/\mathbf{x}}$ are also called *permutation sample spaces*. It is important to note that orbits $\mathcal{X}^n_{/\mathbf{x}}$ associated with data sets $\mathbf{x} \in \mathcal{X}^n$ always contain a finite number of points, as $n$ is finite.

Since, in the null hypothesis and assuming exchangeability, the conditional probability distribution of a generic point $\mathbf{x}' \in \mathcal{X}^n$, for any underlying population distribution $P \in \mathcal{P}$, is $P$-independent, permutation inferences are invariant with respect to $P$ in $H_0$. Some authors, emphasizing this invariance property, prefer to give them the name of invariant tests. However, due to this invariance property, permutation tests are distribution-free and nonparametric. Formally, let $\mathcal{X}^n/\mathbf{x}$ be the orbit associated with the observed vector of data $\mathbf{x}$. The points of $\mathcal{X}^n/\mathbf{x}$ can also be defined as $\mathbf{x}^* : \mathbf{x}^* = \pi\mathbf{x}$ where $\pi$ is a random permutation of indexes $1,2,\ldots,n$.

Define a suitable test statistic $T$ on $\mathcal{X}^n/\mathbf{x}$ for which large values are significant for a right-handed one-sided alternative: The support of $\mathcal{X}^n/\mathbf{x}$ through $T$ is the set $\mathcal{T}$ that consists of $S$ elements (if there are no ties in the given data). Let

$$T^*_{(1)} \leq T^*_{(2)} \leq ,\ldots, \leq T^*_{(S)}$$

be the ordered values of $\mathcal{T}$. Let $T^o$ be the observed value of the test statistic, $T^o = T(\mathbf{x})$. For a chosen attainable significance level $\alpha \in \{1/S, 2/S,\ldots,(S-1)/S\}$, let $k = S(1-\alpha)$. Define a permutation test, the function $\phi^* = \phi(T^*)$ for a one-sided alternative

$$\phi^*(T) = \begin{cases} 1 & \text{if } T^o \geq T^*_{(k)} \\ 0 & \text{if } T^o < T^*_{(k)} \end{cases}.$$

Permutation tests have general good properties such as exactness, unbiasedness and consistency (see Hoeffding, 1952; Pesarin and Salmaso, 2010).

## 1.6 Multivariate permutation tests and nonparametric combination methodology

In this section, we provide details on the construction of multivariate permutation tests via the nonparametric combination approach. Consider, for instance, two multivariate populations within the usual one-way MANOVA layout and the related two-sample multivariate hypothesis testing problem where *p* (possibly dependent) variables are considered.

The one-way MANOVA statistical model (with fixed effects) can be represented as follows:

$$\mathbf{Y}_{ij} = \boldsymbol{\mu}_j + \boldsymbol{\varepsilon}_{ij}, \quad i = 1,...,n_j, \, j = 1,2, \tag{8}$$

where $\boldsymbol{\mu}_j$ is the *p*-dimensional mean effect, $\boldsymbol{\varepsilon}_{ij} \sim \text{IID}(0,\boldsymbol{\Sigma})$ is a *p*-variate random term of experimental errors with zero mean and variance/covariance matrix $\boldsymbol{\Sigma}$. Each univariate component response Y can be of the continuous or binary or ordered categorical moreover the multivariate response can be also mixed (some univariate components are continuous/binary and some other are ordered categorical).

The main difficulties when developing a multivariate hypothesis testing procedure arise because of the underlying dependence structure among variables (or aspects), which is generally unknown and more complex than linear. Moreover, a global answer involving several dependent variables (aspects) is often required, so the question is how to combine the information related to the *p* variables (aspects) into one global test.

In order to better explain the proposed approach let us denote an $n \times p$, $n = n_1 + n_2$, data set with $\mathbf{Y}$:

$$\mathbf{Y} = \begin{bmatrix} \mathbf{Y}_1 \\ \mathbf{Y}_2 \end{bmatrix} = [Y_1, Y_2, ..., Y_p] = \begin{bmatrix} y_{11} & y_{12} & \cdots & y_{1p} \\ y_{21} & y_{22} & \cdots & y_{2p} \\ \cdots & \cdots & \cdots & \cdots \\ y_{n1} & y_{n2} & \cdots & y_{np} \end{bmatrix},$$

where $\mathbf{Y}_1$ and $\mathbf{Y}_2$ are the $n_1 \times p$ and the $n_2 \times p$ samples drawn from the first and second population respectively. In the framework of NonParametric Combination (NPC) of Dependent Permutation Tests we suppose that, if the global null hypothesis $H_0$: $\boldsymbol{\mu}_1 = \boldsymbol{\mu}_2$ of equality of the two populations is true, the hypothesis of exchangeability of random errors holds. Hence, the following set of mild conditions should be jointly satisfied:

a) we suppose that for $\mathbf{Y}=[\mathbf{Y}_1,\mathbf{Y}_2]$ an appropriate probabilistic $p$-dimensional distribution structure $P$ exists, $P_j \in \mathcal{P}$, $j=1,2$, belonging to a (possibly non-specified) family $\mathcal{P}$ of non-degenerate probability distributions;

b) the null hypothesis $H_0$ states the equality in distribution of the multivariate distribution of the $p$ variables in all 2 groups:

$$H_0 : [P_1 = P_2] = \left[ \mathbf{Y}_1 \stackrel{d}{=} \mathbf{Y}_2 \right].$$

Null hypothesis $H_0$ implies the exchangeability of the individual data vector with respect to the 2 groups. Moreover, according to Roy's Union Intersection Criterion (1953), $H_0$ is supposed to be properly decomposed into $p$ sub-hypotheses $H_{0k}$, $k=1,\ldots,p$, each appropriate for partial (univariate) aspects, thus $H_0$ (multivariate) is true if all the $H_{0k}$ (univariate) are jointly true:

$$H_0 : \left[ \bigcap_{k=1}^{p} Y_{1k} \stackrel{d}{=} Y_{2k} \right] = \left[ \bigcap_{k=1}^{p} H_{0k} \right].$$

$H_0$ is called the *global* or *overall null hypothesis*, and $H_{0k}$, $k=1,\ldots,p$, are called the *partial null hypotheses*.

c) The alternative hypothesis $H_1$ is represented by the union of partial $H_{1k}$ sub-alternatives:

$$H_1 : \left[ \bigcup_{k=1}^{p} H_{1k} \right],$$

so that $H_1$ is true if at least one of sub-alternatives is true.

In this context, $H_1$ is called the *global* or *overall alternative*, and $H_{1k}$, $k=1,...,p$, are called the *partial alternatives.*

d) let $\mathbf{T}=\mathbf{T}(\mathbf{Y})$ represent a *p*-dimensional vector of test statistics, $p \geq 1$, whose components $T_k$, $k=1,...,p$, represent the partial univariate and non-degenerate *partial test* appropriate for testing the sub-hypothesis $H_{0k}$ against $H_{1k}$. Without loss of generality, all partial tests are assumed to be marginally unbiased, consistent and significant for large values (for more details and proofs see Pesarin and Salmaso, 2010).

At this point, in order to test the global null hypothesis $H_0$ and the *p* univariate hypotheses $H_{0k}$, the key idea comes from the partial (univariate) tests which are focused on *k*-th partial aspects, and then, combining them with an appropriate combining function, firstly to test $H_{0k}$, $k=1,...,p$, and finally to test the global (multivariate) test which is referred to as the global null hypothesis $H_0$.

However, we should observe that in most real problems when the sample sizes are large enough, there is a clash over the problem of computational difficulties in calculating the conditional permutation space. This means it is not possible to calculate the exact *p*-value of observed statistic $T_{k0}$. This is brilliantly overcome by using the CMCP (Conditional Monte Carlo Procedure). The CMCP on the pooled data set $\mathbf{Y}$ is a random simulation of all possible permutations of the same data under $H_0$ (for more details refer to Pesarin and Salmaso, 2010). Hence, in order to obtain an estimate of the permutation distribution under $H_0$ of all test statistics, a CMCP can be used. Every resampling without replacement $\mathbf{Y}^*$ from the data set $\mathbf{Y}$ actually consists of a random attribution of the individual block data vectors to the C treatments. In every $\mathbf{Y}_r^*$ resampling, $r=1,...,B$, the *k* partial tests are calculated to obtain the set of values $[\mathbf{T}_{ir}^*=\mathbf{T}(\mathbf{Y}_{ir}^*)$, $i=1,..,k$; $r=1,...,B]$, from the B independent random resamplings. It should be emphasized that CMCP only considers permutations of individual data vectors, so that all underlying dependence relations which are present in the component variables are preserved. From this point of view, the CMCP is essentially a multivariate procedure.

Without loss of generality, let us suppose that all partial tests are significant for large values. More formally, the steps of the CMC procedure are described as follows:

1. calculate the $p$-dimensional vectors of statistics, each one related to the corresponding partial tests from the observed data:

$$\mathbf{T}^{obs}_{p \times 1} = \mathbf{T}(\mathbf{Y}) = \left[ T_k^{obs} = T_k(\mathbf{Y}), k = 1, \ldots, p \right],$$

2. calculate the same vectors of statistics for the permuted data:

$$\mathbf{T}_b^* = \mathbf{T}(\mathbf{Y}_b^*) = \left[ T_{bk}^* = T_k(\mathbf{Y}_b^*), k = 1, \ldots, p \right],$$

3. repeat the previous step $B$ times independently. We denote with $\{\mathbf{T}_b^*, b=1,\ldots,B\}$ the resulting sets from the $B$ conditional resamplings. Each element represents a random sample from the $p$-variate permutation c.d.f. $F_T(z|\mathbf{Y})$ of the test vector $\mathbf{T}(\mathbf{Y})$.

The resulting estimates are:

$$\hat{F}_T(\mathbf{z} \mid \mathbf{Y}) = \left[ \frac{1}{2} + \sum_{b=1}^{B} \mathbb{I}(\mathbf{T}_b^* \leq \mathbf{z}) \right] \bigg/ (B+1), \forall \mathbf{z} \in \mathbb{R}^p,$$

$$\hat{L}_{T_k}(z \mid \mathbf{Y}) = \left[ \frac{1}{2} + \sum_{b=1}^{B} \mathbb{I}(\mathbf{T}_{bk}^* \geq z) \right] \bigg/ (B+1), \forall z \in \mathbb{R}^1,$$

$$\hat{\lambda}_k = \hat{L}_{T_k}(T_k^{obs} \mid \mathbf{Y}) = \left[ \frac{1}{2} + \sum_{b=1}^{B} \mathbb{I}(\mathbf{T}_{bk}^* \geq T_k^{obs}) \right] \bigg/ (B+1), k = 1, \ldots, p,$$

where $\mathbb{I}(.)$ is the indicating function and where with respect to the traditional EDF estimators, 1/2 and 1 have been added respectively to the numerators and denominators in order to obtain estimated values in the open interval (0,1), so that transformations by inverse CDF of continuous distributions are continuous and so are always well defined.

Hence, if $\hat{\lambda}_k < \alpha$ the null hypothesis corresponding to the *k*-th variable ($H_{0k}$) is rejected at significance level equal to $\alpha$ (adjusted for multiplicity).

Moreover, choice of partial tests has to provide that:

(i) all partial tests $T_k$ are marginally unbiased, formally:

$$\mathbb{P}\{T_k \geq z \mid \mathbf{Y}, H_{0k}\} \leq \mathbb{P}\{T_k \geq z \mid \mathbf{Y}, H_{1k}\}, \forall z \in \mathbb{R}^1,$$

(ii) at least one partial tests is consistent, i.e.

$$\mathbb{P}\{T_k \geq T_{k\alpha} \mid H_{1k}\} \to 1, \forall \alpha > 0 \text{ as } n \to \infty,$$

where $T_{k\alpha}$ is a finite $\alpha$-level for $T_k$.

Let us now consider a suitable continuous non-decreasing real function, $\phi: (0,1)^p \to \mathbb{R}^1$, that applied to the *p*-values of partial tests T defines the second order global (multivariate) test T″:

$$T'' = \phi(\lambda_1, \ldots, \lambda_p),$$

provided that the following conditions hold:

- $\phi$ is non-increasing in each argument: $\phi(\ldots, \lambda_k, \ldots) \geq \phi(\ldots, \lambda'_k, \ldots)$, if $\lambda_k \leq \lambda'_k$, $k \in \{1, \ldots, p\}$;

- $\phi$ attains its supremum value $\bar{\phi}$, possibly not finite, even when only one argument attains zero: $\phi(\ldots, \lambda_k, \ldots) \to \bar{\phi}$ if $\lambda_k \to 0$, $k \in \{1, \ldots, p\}$;

- $\phi$ attains its infimum value $\underline{\phi}$, possibly not finite, when all its arguments attain one: $\phi(\ldots, \lambda_k, \ldots) \to \underline{\phi}$ if $\lambda_k \to 1$, $k = 1, \ldots, p$;

- $\forall \alpha > 0$, the acceptance region is bounded: $\underline{\phi} < T''_{\alpha/2} < T'' < T''_{1-\alpha/2} < \bar{\phi}$.

Frequently used combining function are:

- Fisher combination: $\phi_F = -2\sum_k \log(\lambda_k)$;

- Tippet combination: $\phi_T = \max_{1 \leq k \leq p}(1 - \lambda_k)$;

- Direct combination: $\phi_D = \sum_k T_k$;

- Liptak combination: $\phi_L = \sum_k \Phi^{-1}(1-\lambda_k)$;

where $k = 1,\ldots,p$ and $\Phi$ is the standard normal c.d.f.

It can be seen that under the global null hypothesis the CMC procedure allows for a consistent estimation of the permutation distributions, marginal, multivariate and combined, of the $k$ partial tests. Usually, Fisher's combination function is considered (Pesarin and Salmaso, 2010), mainly for its finite and asymptotic good properties. Of course, it would be also possible to take into consideration any other combining function (Lancaster, Mahalanobis, etc.; see Folks, 1984; Pesarin and Salmaso, 2010). In the stated conditions the combined test is also unbiased and consistent. For a detailed description we refer to Pesarin and Salmaso (2010).

## 1.7 Hypothesis testing for curves comparison

Important inferential problems usually occur when the data of interest are a collection of scalars or vectors which can be viewed as samples drawn from population of curves or trajectories. Thank to modern technologies such kind of data are more and more frequently observed in many different areas and contexts. Let us think for examples to spectrometric curves, radar waveforms, gene expressions, etc. From the statistical point of view this type of data can be viewed as either longitudinal data from repeated measures on the same units/subjects (Diggle et al., 2004) or observations, called functional data, lying to infinite dimensional spaces commonly called functional spaces (Ferraty and Vieu, 2006). Functional data are characterized in general by data collected regularly at a high frequency, while longitudinal data is usually more sparse in time and collected at irregular intervals (Rice, 2004). Consequently, functional data analysis focuses more on dimension reduction. In addition, longitudinal data analysis is more model-based and inferential, while functional data analysis has a more exploratory and nonparametric point of view with a focus on describing the data (with principal components, smoothing, etc.).

### 1.7.1 Literature review on testing for curves comparison

Until recently, functional data analysis (FDA) and longitudinal data analysis (LDA) have been viewed as distinct enterprises. As of a 2004 special issue of Statistica Sinica, it is seen that endeavors have been made to reconcile the two lines of research. It is worth noting that although functional data analysis and longitudinal data analysis are both devoted to analyzing curves/trajectories on the same subjects, FDA and LDA are also intrinsically different (Davidian et al., 2004). Longitudinal data are involved in follow up studies (common on biomedical sciences) which usually require several (few) measurements of the variables of interest for each individual along the period of study. They are often treated by multivariate parametric techniques which study the variation among the means along the time controlled by a number of covariates. In contrast, functional data are frequently recorded by mechanical instruments (more common in engineering and physical sciences although also in an increasing number of biomedical problems) which collect many repeated measurements per subject. Its basic units of study are complex objects such as curves (commonly), images or shapes (information along the time of the same individual is jointly considered). Conceptually, functional data can be considered sample paths of a continuous stochastic process (Valderrama, 2007) where the usual focus is studying the covariance structure. In addition, the infinite dimensional structure of the functional data makes that the links with standard nonparametric statistics (in particular with smoothing techniques) particularly strong (González-Manteiga and Vieu, 2007). Despite these differences, which involve mainly the viewpoints and ways of thinking about the data of both fields, Zhao et al. (2004) connected them illustrating the ideas in the context of a gene expression study example, introduced LDA to the FDA viewpoint.

As pointed out by Sirski (2012), despite the fact that the comparison between the curves is not only of methodological interest but has important practical implications especially for technological and biomedical applications, much of the literature in the field of FDA is concerned with describing the data (with principal components, smoothing, etc.) as opposed to formal hypothesis testing, including assessing statistical significance between groups of curves. Even if the emphasis is on describing curves, the FDA literature proposes some solutions for the testing problem of comparing population of curves. Beyond the FDA proposals several other testing solutions can be also identified in the field of LDA. From a critical review on hypothesis testing solutions for curves comparison it appears that in the literature there are basically two main approaches which may be briefly described as basis function approximation solutions and overall tests. The first class of procedures are concerned with testing on the equality of the coefficients from a basis function approximation while the second one is aimed at developing a global test which compares the population of curves using a suitable test statistic defined in the whole domain of the functional response. Similar to other testing problems, both the FDA and LDA literature contain parametric and nonparameteric solutions for the two approaches.

Let us begin our review with the basis function approximations. The rationale behind this approach for hypothesis testing on curves is based on the principle of preliminarily transforming and reducing data in order to reduce the dimensionality of the problem by using a suitable transformation such as Fourier, wavelet, principal component analysis, etc. then testing for the equality among groups of the related coefficients. This principle applies both for the parametric and for the nonparametric framework. From a parametric point of view, a class of basis function solutions may be referred to for the thresholding methods for testing problems. Fan and Lin (1998) developed some adaptive Neyman tests for curve data from stationary Gaussian processes using orthogonal transformations such as Fourier and wavelet to preprocess the data and compress signals (Fan, 1996). In case the comparison involves more than two set of curves, authors call the proposed solutions as HANOVA, i.e. the analysis of variance when the dimensionality is high.

Within a nonparametric framework for curve comparisons using basis function approximations, Eubank (2000) proves that among the different ways of combining the coefficients into a test statistic, the L2 norm, a simple sum of the squared coefficients, is asymptotically equivalent to the uniformly most powerful test when the grid size goes to infinity. Zhang and Chen (2007) propose a L2-norm-based global test statistic based on the local polynomial kernel smoothing technique demonstrating how to reconstruct individual functions from a discrete functional data set using local polynomial smoothing. They show that, under some mild conditions, the effects of substitutions of the functions with their local polynomial reconstructions can be ignored asymptotically.

In a generalized nonparametric regression framework Behseta and Kass (2005) propose two methods referred to as the so-called Bayesian adaptive regression splines (BARS). The first method uses Bayes factors, and the second method uses a modified Hotelling-type test. Behseta et al. (2007) extend the application of BARS to the likelihood ratio tests based on the asymptotic distribution of BARS fits where the reference distribution of the test statistics are derived asymptotically or via bootstrap.

Let us move the focus on the overall test solutions for curves comparison. In this connection and with respect to the parametric proposals, a class of solutions has been inspired by the attempt to extend the ANOVA techniques in the context of functional data. In their classical text book reference for FDA, Ramsay and Silverman (2005) suggest calculating the F-test statistic at each time point, but do not address how to deal with the resulting statistics. Shen and Faraway (2004) suggest a functional F test for comparing nested functional linear models; the null distribution of which is derived by approximation. The reference model for the Shen and Faraway solution is the so-called varying-coefficient models with time fixed covariates which is a special case for the general varying-coefficient model with both time varying or time fixed covariates (Wu and Yu, 2002). Shen and Faraway applies their functional F test to some data from ergonomics and investigate its nominal size and power by a simulation study including some competing tests such as bootstrap methods as described in Faraway (1997), the traditional multivariate log likelihood ratio test on raw data and a test based on a B-spline basis function representation. Their simulations show that the power of the tests is data-dependent and argue that the F test has the benefit of avoiding the risk that the other two comparison tests have with being influenced by 'unimportant directions of variation'. A simulation study on the comparison between the functional F test with the multivariate likelihood ratio and B-spline–based tests is presented also in Shen and Xu (2007) where results confirms that the better performance of the functional F test is setting-dependent. In the same direction Yang et al. (2007) illustrates how to apply the Shen and Faraway functional F test to the setting of longitudinal data. They conduct a simulation study to investigate the statistical power for the F test compared with the Wilks' likelihood ratio test and the linear mixed-effects model using AIC. Results confirms the general good behavior of the functional F test and its power is proved one more time to be setting-dependent, specifically the covariance structure of the error process affects the power of the test. Xu et al. (2011) propose a quasi F-test for functional linear models with functional covariates and outcomes which is an extension of the Shen and Faraway's F-statistic. From simulations to study the size and the power of the quasi F-test when comparing it with a linear mixed effects model approach, Xu et al. observe that their proposed test is more powerful than the linear mixed effects method only in case of large effects. Zhang (2011) propose a new functional F-type test in order to reduce the bias in the approximate null distribution of the Shen and Faraway's F-test. In a simulation study Zhang demonstrates that the bias-reduced method and the naive method perform similarly when the data are highly or moderately correlated, but the former outperforms the latter significantly when the data are nearly uncorrelated. Always in the spirit of ANOVA techniques in the context of functional data, Cuevas et al. (2004) suggest an asymptotic version of the well-known ANOVA F-test in the case of functional data. To overcome some difficulties in practice with handling the asymptotic distribution of the test statistic they propose a numerical Monte Carlo approach. In the same line with the proposal of Cuevas et al., Martínez-Camblor and Corral (2011) suggest a generalization of the classical ANOVA F-ratio for repeated measures to the functional setting. Both the parametric and the nonparametric approaches are considered to derive the asymptotic and the resampling distribution of the test statistic. Within a nonparametric framework, asymptotic distributions of ANOVA-type test statistics are presented in Wang and Akritas (2010; 2011). The test statistics are in the form of a difference of two quadratic forms and have a limiting Chi-square and normal distribution.

For nonparametric methods whose goal is to develop a global test suitable for the problem of comparing curves, there are two main approaches in the literature: permutation-based solutions and nonparametric regression solutions. Let us first consider the permutation approach, which seems very popular for the testing problem at hand probably due to the flexibility of permutation solutions in handling complex testing problems, especially when the asymptotic distributions are difficult to derive and/or the parametric assumptions are hard to justify (Pesarin and Salmaso, 2010). From the notion of similarity between two curves and in order to test the null hypothesis of no difference, Munoz Maldonado et al. (2002) suggest three permutation statistics, i.e. pooled-mean, pooled-variance and the ratio between them. A similar solution is proposed in the work of Sturino et al. (2010); after quantifying the distance between mean or median curves from two treatments, apply a permutation test using as test statistic, either the difference of the means (or medians) or the difference of the areas between mean curves. They validate via simulations the proposed solutions, including also an additional permutation statistic employing functional principal component analysis. Using a different strategy based on pairwise differences between individual curves, Sirski (2012) proposes a set of permutation statistics and compares them by simulation with a collection of other permutation tests proposed by literature, a test based on the functional principal components scores (Sturino et al., 2010), the adaptive Neyman test (Fan and Lin, 1998) and the functional F test (Shen and Faraway, 2004). Sirski's simulation results suggest that the solution with the best power performance is the test based on the mean of the pairwise L1-norms, while the worst solution is the functional F test which has a poor performance in case of non-normal errors and of small sample sizes. The joint use of permutation tests and the nonparametric combination methodology allows Pesarin and Salmaso (2010) to propose a global test they call time-to-time permutation test able to properly combine in a suitable global test the set of the dependent univariate permutation tests performed in all observation points. A permutation solution inspired from a multiple comparison procedure is proposed by Cox and Lee (2008) where a set of pointwise t-tests are applied on smooth functional data. Similarly, Ramsay et al. (2009) address the issue of hypothesis testing proposing a permutation approach based on the absolute value of the test statistic similar in form to the t-test statistic at each point in time. They also propose a functional F statistic and use a permutation test based on the maximum F value.

Alternative nonparametric solutions are proposed in the literature within the nonparametric regression framework. In this connection Cardot et al. (2007) suggest a permutation approach to check if a real covariate has a significant effect on a functional response in a regression setting using two test statistics, i.e. an adapted F-statistic and a statistic based on the kernel smoother applied to the residuals. Zhang and Lin (2003) propose a solution for testing the equivalence of two nonparametric functions in semiparametric additive mixed models for correlated non-Gaussian data. This test extends the previous work of Zhang et al. (2000). Neumeyer and Dette (2003) propose a new test for the comparison of two regression curves that is based on a difference of two marked empirical processes based on residuals which is applicable in the case of different design points and heteroscedasticity. Finally, from a functional data analysis and nonparametric regression perspective Hall and van Keilegom (2007) suggest a Cramer-von Mises type test and take up the issue of studying how the data pre-processing interferes with the performance of two-sample statistical tests.

### 1.7.2 The time-to-time permutation solution

In order to formalize the problem of comparing two or more population of curves within the permutation and combination methodology let us note that functional data can also be interpreted as discrete or discretized stochastic processes for which at most a countable set of data points are observed. In this way an observed curve is nothing more than a set of *repeated measures* in which each subject/unit (a given curve) is observed on a finite or at most a countable number of occasions so that successive responses are dependent (Pesarin and Salmaso, 2010). In practice, responses of one unit may be viewed as obtained by a discrete or discretized stochastic process.

With reference to each specific subject, repeated observations are also called the *response profiles*, and may be viewed as a multivariate variable. In the context of the repeated measurements designs, an existing permutation solution has already been proposed by Pesarin and Salmaso (2010) essentially employ the method of nonparametric combination of dependent permutation tests, each obtained by a partial analysis on data observed on the same ordered occasion (so-called *time-to-time analysis*).

Without loss of generality, we discuss general problems which can be referred to in terms of a one-way MANOVA layout for response profiles. Hence, we refer to testing problems for treatment effects when:

a) measurements are typically repeated a number of times on the same units;

b) units are partitioned into $C$ groups or samples, there being $C$ levels of a treatment;

c) the hypotheses being tested aim to verify whether the observed profiles do or do not depend on treatment levels;

d) it is presumed that responses may depend on time, space, etc. and that related effects are not of primary interest.

For simplicity, from here onwards we refer to time occasions of observation, where time means any sequentially ordered entity including: space, lexicographic ordering, etc.

Let us assume that the permutation testing principle holds, in particular, in the null hypothesis, in which treatment does not induce differences with respect to levels, we assume that the individual response profiles are exchangeable with respect to groups. To be more specific, let us refer to a problem in which $n$ units are partitioned into $C$ groups and a univariate variable $X$ is observed. Groups are of size $n_j \geq 2$, $j=1,...,C$, with $n=\sum_j n_j$. Units belonging to the $j$th group are presumed to receive a treatment at the *j*th level. All units are observed at *N* fixed ordered occasions $\tau_1,..., \tau_N$, where *N* is an integer. For simplicity, we refer to time occasions by using *t* to mean $\tau_t$, $t=1,...,N$. Hence, for each unit, we observe the discrete or discretized profile of a stochastic process, and profiles related to different units are assumed to be stochastically independent. Thus, within the hypothesis that treatment levels have no effect on response distributions, profiles are exchangeable with respect to groups.

Let us refer to a univariate stochastic time model with additive effects, covering a number of practical situations. Extensions of the proposed solution to multivariate response profiles are generally straightforward.

The symbol $\mathbf{X}=\{X_{ji}(t), i=1,...,n_j, j=1,...,C, t=1,...,N\}$ indicates that the whole set of observed data is organized as a two-way layout of univariate observations. Alternatively, especially when effects due to time are not of primary interest, $\mathbf{X}$ may be organized as a one-way layout of profiles, $\mathbf{X}=\{\mathbf{X}_{ji}(t), i=1,...,n_j, j=1,...,C\}$, where $\mathbf{X}=\{X_{ji}(t), t=1,...,N\}$ indicates the *ji*th observed profile.

The general additive response model referred to this context is

$$X_{ji}(t) = \mu + \eta_j(t) + \varepsilon_{ji}(t),$$

where $\varepsilon_{ji}(t) = \Delta_{ji}(t) + \sigma(\eta_j(t)) \cdot Z_{ji}(t)$, $i=1,...,n_j$, $j=1,...,C$, $t=1,...,N$. In this model, $Z_{ji}(t)$ are generally non-Gaussian error terms distributed as a stationary stochastic process with null mean and unknown distribution $P_\mathbf{Z}$ (i.e. a generic white noise process); these error terms are assumed to be exchangeable with respect to units and treatment levels but, of course, not independent of time. Moreover, $\mu$ is a population constant; coefficients $\eta_j(t)$ represent the *main treatment effects* and may depend on time through any kind of function, but are independent of units; quantities $\Delta_{ji}(t)$ represent the so-called *individual effects*; and $\sigma(\eta_j(t))$ are time-varying scale coefficients which may depend, through monotonic functions, on main treatment effects $\eta_j$ provided that the resulting CDFs are pairwise ordered so that they do not cross each other, as in $X_j(t) \stackrel{d}{<}$ (or $\stackrel{d}{>}$) $X_r(t)$, $t=1,...,N$ and $j \neq r=1,...,C$. When $\Delta_{ji}(t)$ are stochastic, we assume that they have null mean values and distributions which may depend on main effects, units and treatment levels. Hence, random $\Delta_{ji}(t)$ are determinations of an unobservable stochastic process or, equivalently, of a $k$-dimensional variable $\Delta = \{\Delta(t), t=1,...,N\}$. In this context, we assume that $\Delta_j \sim \mathcal{D}_k\{\mathbf{0}, \beta(\eta_j)\}$, where $\mathcal{D}_k$ is any unspecified distribution with null mean vector and unknown dispersion matrix $\beta$, indicating how unit effects vary with respect to main effects $\eta_j = \{\eta_j(t), t=1,...,N\}$. Regarding the dispersion matrix $\beta$, we assume that the resulting treatment effects are pairwise stochastically ordered, as in $\Delta_j(t) \stackrel{d}{<} (\text{or} \stackrel{d}{>}) \Delta_r(t)$, $t=1,...,N$ and $j \neq r=1,...,C$. Moreover, we assume that the underlying bivariate stochastic processes $\{\Delta_{ji}(t), \sigma(\eta_j(t)) \cdot Z_{ji}(t)), t=1,...,N\}$ of individual stochastic effects and error terms, in the null hypothesis, are exchangeable with respect to groups. This property is easily justified when subjects are randomized to treatments.

This setting is consistent with a general form of dependent random effects fitting a very large number of processes that are useful in most practical situations. In particular, it may interpret a number of the so-called *growth processes*. Of course, when $\beta = \mathbf{0}$ with probability one for all $t$, the resulting model has fixed effects.

In order to appreciate the inherent difficulties in statistical analysis of real problems when repeated observations are involved, see, for example, Diggle et al. (2002). In particular, when dispersion matrices $\Sigma$ and $\beta$ have no known simple structure, the underlying model may not be identifiable and thus no parametric inference is possible. Also, when $N \geq n$, the problem cannot admit any parametric solution (see Blair et al., 1994, in which heuristic solutions are suggested under normality of errors $\mathbf{Z}$ and for fixed effects).

Among the many possible specifications of models for individual effects, one of these assumes that terms $\Delta_{ji}(t)$ behave according to an AR(1) process:

$$\Delta_{ji}(0) = 0; \quad \Delta_{ji}(t) = \chi(t) \cdot \Delta_{ji}(t-1) + \beta(\eta_j(t)) \cdot W_{ji}(t),$$

$i=1,...,n_j$, $j=1,...,C$, $t=1,...,N$, where $W_{ji}(t)$ represent random contributions interpreting deviates of individual behaviour; $\gamma_j(t)$ are autoregressive parameters which are assumed to be independent of treatment levels and units, but not time; $\beta(\eta_j(t))$, $t=1,...,N$, are time-varying scale coefficients of autoregressive parameters, which may depend on the main effects. By assumption, the terms $W_{ji}(t)$ have null mean value, unspecified distributions, and are possibly time-dependent, so that they may behave as a stationary stochastic process.

A simplification of the previous model considers a regression-type form such as

$$\Delta_{ji}(t) = \gamma_j(t) + \beta(t) \cdot W_{ji}(t), \ i=1,...,n_j, \ j=1,...,C, \ t=1,...,N.$$

Note that many other models of dependence errors might be taken into consideration, including situations where matrices $\Sigma$ and $\beta$ are both full.

Within the above presented layout, the hypotheses of interest we wish to test are

$$H_0 : \left\{ \mathbf{X}_1 \stackrel{d}{=} ... \stackrel{d}{=} \mathbf{X}_C \right\} = \left\{ X_1(t) \stackrel{d}{=} ... \stackrel{d}{=} X_C(t), t=1,...,N \right\}$$

$$\left\{ \bigcap_{t=1}^{N} \left[ X_1(t) \stackrel{d}{=} ... \stackrel{d}{=} X_C(t) \right] \right\} = \left\{ \bigcap_{t=1}^{N} H_{0t} \right\},$$

against $H_1: \{\bigcup_t [H_{0t} \text{ is not true}]\} = \{\bigcup_t H_{1t}\}$, in which a decomposition of the global hypotheses into $N$ sub-hypotheses according to time is highlighted. This decomposition corresponds to the so-called *time-to-time* analysis for which an existing permutation solution has already been proposed by Pesarin and Salmaso (2010).

Note that by decomposition into $N$ partial sub-hypotheses, each sub-problem is reduced to a one-way ANOVA. Also note that, from this point of view, the associated two-way ANOVA, in which effects due to time are not of interest, becomes equivalent to a one-way MANOVA.

In the given conditions, $N$ partial permutation tests $T_t^* = \sum_j n_j \cdot \left( \bar{X}_j^* \right)^2$, where $\bar{X}_j^* = \sum_i X_{ji}^*(t) / n_j$, $t=1,...,N$ are appropriate for time-to-time sub-hypotheses $H_{0t}$ against $H_{1t}$.

Thus, in order to achieve a global complete solution for $H_0$ against $H_1$, we must combine all these partial tests. Of course, due to the complexity of the problem and to the unknown *N*-dimensional distribution of ($T_1,..., T_N$) (see Diggle et al., 2002), we are generally unable to evaluate all dependence relations among partial tests directly from **X**. Therefore, this combination should be nonparametric and may be obtained through any combining function $\phi \in \mathcal{C}$. Of course, when the underlying model is not identifiable, and so some or all of the coefficients cannot be estimated, this NPC becomes unavoidable. Moreover, when all observations come from only one type of variable (continuous, discrete, nominal, ordered categorical) and thus partial tests are homogeneous, a direct combination of standardized partial tests, such as $T_t^* = \sum_j n_j \cdot \left[\bar{X}_j^*(t) - \bar{X}_\bullet(t)\right]^2 \Big/ \sum_{ji} \left[X_j^*(t) - \bar{X}_j^*(t)\right]^2$, may be appropriate especially when *N* is large.

## 1.8 Some properties of univariate and multivariate permutation tests

In this section we present two important properties of univariate and multivariate permutation tests, specifically the equivalence of permutation statistics and the so-called finite sample consistency.

### 1.8.1 Equivalence of permutation statistics

The concept of permutationally equivalent statistics is useful in simplifying computations and sometimes in facilitating the establishment of the asymptotic equivalence of permutation solutions with respect to some of their parametric counterparts.

Let **Y** be the given data set from a response variable *Y*, so that Y belongs to the sample space $\Omega$. We have the following definition of the permutation equivalence of two statistics:

**Definition.** Two statistics $T_1$ and $T_2$, both mapping $\Omega$ into $\Re^1$, are said to be permutationally equivalent when, for all points $\mathbf{Y} \in \Omega$ and $\mathbf{Y}^* \in \Omega_{/\mathbf{Y}}$, the relationship $\{T_1(\mathbf{Y}^*) \leq T_1(\mathbf{Y})\}$ is true if and only if $\{T_2(\mathbf{Y}^*) \leq T_2(\mathbf{Y})\}$ is true, where $\mathbf{Y}^*$ indicates any permutation of **Y** and $\Omega_{/\mathbf{Y}}$ indicates the associated permutation sample space. This permutation equivalence relation is indicated by $T_1 \approx T_2$.

With reference to this definition we have the following theorem and corollaries (Pesarin, 2001).

**Theorem A.1** If between the two statistics $T_1$ and $T_2$ there is a one-to-one increasing relationship, then they are permutationally equivalent and $\Pr\{T_1(\mathbf{Y}^*) \leq T_1(\mathbf{Y})|\mathbf{Y}\}=\Pr\{T_2(\mathbf{Y}^*) \leq T_2(\mathbf{Y})\}|\mathbf{Y}\}$, where these probabilities are evaluated with respect to conditional distribution $P_{|\mathbf{Y}}$ induced by the sampling experiment and defined on the permutation measurable space $(\Omega_{/\mathbf{Y}}, B_{/\mathbf{Y}})$.

**Corollary A.1** If $T_1$ and $T_2$ are related by an increasing one-to-one relationship with probability one, then they are permutationally equivalent with probability one, where this probability is measured in terms of population distribution $P$.

**Corollary A.2** If $T_1$ and $T_2$ are related by a decreasing one-to-one relationship, then they are permutationally equivalent in the sense that $\{T_1(\mathbf{Y}^*) \leq T_1(\mathbf{Y})\} \leftrightarrow \{T_2(\mathbf{Y}^*) \geq T_2(\mathbf{Y})\}|\mathbf{Y}\}$ for all $\mathbf{Y} \in \Omega$ and $\mathbf{Y}^* \in \Omega_{/\mathbf{Y}}$.

**Corollary A.3** The permutation equivalence relation is reflexive: $T_1 \approx T_1$.

**Corollary A.4** The permutation equivalence relation is transitive: if $T_1 \approx T_2$ and $T_2 \approx T_3$, then $T_1 \approx T_3$.

### 1.8.2 The finite sample consistency

A quite important problem usually occurs in some multidimensional applications when sample sizes are fixed and the number of variables which are to be analyzed is much larger than sample sizes (Blair et al., 1994; Goggin, 1986). In Pesarin (2001) it is shown that, under very mild conditions, the power function of permutation tests based on both associative and non-associative statistics monotonically increases as the related standardized noncentrality functional increases. This is true also for multivariate situations. In particular, for any added variable the power does not decrease if this variable makes larger standardized noncentrality (*finite-sample consistency*). These results confirm and extend those presented by Blair et al. (1994) and Pesarin and Salmaso (2010b). In particular, Blair et al. (1994) presents an exhaustive power simulation study comparing permutation tests and Hotelling's $T^2$ test when the number of variables increases with respect to fixed sample sizes and shows a better behaviour of the permutation tests.

In order to present the finite-sample consistency we refer to one-sided two-sample designs for non-negative alternatives. Extensions to non-positive and/or two-sided alternatives, and multisample designs are straightforward. Let $\mathbf{Y}_j=\{Y_{ij};\ i = 1,...,n_j\} \in \mathcal{Y}^{n_j}$ the independent and identically distributed (i.i.d.) sample data of size $n_j$ from $(Y, \mathcal{Y}, P_j \in \mathcal{P})$, $j = 1,2$, where $Y$ is the variable of interest taking values in the sample space $\mathcal{Y}$ according to the distribution $P_j$. A notation for data sets with independent samples is $\mathbf{Y} = \{Y_{11},...,Y_{1n_1}, Y_{21},...,Y_{2n_2}\} \in \mathcal{Y}^n$, where $n = n_1 + n_2$. To denote data sets in the permutation context it is sometimes convenient to use the unit-by-unit representation: $\mathbf{Y}=\{Y_i;\ i=1,...,n;\ n_1,n_2\}$, where it is intended that first $n_1$ data in the list belong to first sample and the rest to the second. In practice, denoting by $(u_1^*,...,u_n^*)$ a permutation of $(1,...,n)$, $\mathbf{Y}^*=\{Y_i^* = Y(u_i^*, i=1,...,n;\ n_1,n_2\}$ is the related permutation of $\mathbf{Y}$, so that $\mathbf{Y}_1^* =\{Y_{1i}^*=Y(u_i^*), i=1,...,n_1\}$ and $\mathbf{Y}_2^* =\{Y_{2i}^*=Y(u_i^*), i= n_1+1,...,n_2\}$ are the two permuted samples, respectively.

Here we discuss testing problems for stochastic dominance alternatives as are generated by treatments with non-negative shift effects $\boldsymbol{\delta}$. In particular, the alternative assumes that treatments produce an effect $\boldsymbol{\delta}$ so that $\boldsymbol{\delta} > \mathbf{0}$. Thus, the hypotheses are

$$H_0: \left[ \mathbf{Y}_1 \stackrel{d}{=} \mathbf{Y}_2 \stackrel{d}{=} \mathbf{Y} \right] \equiv \left[ P_1 = P_2 \right] \text{ vs. } H_1: \left[ \mathbf{Y}_1 + \boldsymbol{\delta} \stackrel{d}{>} \mathbf{Y}_2 \right].$$

Note that under $H_0$ data of two samples are exchangeable, in accordance with the notion that subjects are randomized to treatments. Since effects $\boldsymbol{\delta}$ may depend on null responses $Y_1$, stochastic dominance $(Y_1+\delta) \overset{d}{>} Y_2 = Y$ is compatible with non-homoscedasticities in the alternative. Thus, the null hypothesis may also be written as $H_0:[\boldsymbol{\delta}=0]$. Sometimes, to emphasize the role of effects we use

$$\mathbf{Y}(\boldsymbol{\delta})=\{Y_{11}+\delta,\ldots,Y_{1n_1}+\delta,Y_{21},\ldots,Y_{2n_2}\}$$

to denote data sets, and so $\mathbf{Y}(0)$ denotes data in $H_0$. In this context, it is also worth noting that observed variable $Y$, sample space $\mathcal{Y}$, and effect $\boldsymbol{\delta}$ are $p$-dimensional, with $p \geq 1$.

In this paper we consider test statistics based on comparison of sampling indicators like $T^*(\boldsymbol{\delta})=S(\mathbf{Y}_1^*(\boldsymbol{\delta}))-S(\mathbf{Y}_2^*))$, where $S(\mathbf{Y}_j^*(\boldsymbol{\delta}))$: $\mathcal{Y}^{nj} \to \mathcal{R}^1$ is any symmetric function, i.e. invariant with respect to rearrangements of entry arguments. These kind of statistics include associative forms like

$$T^*(\boldsymbol{\delta})=1/n_1 \sum_i \varphi[\mathbf{Y}_{1j}^*(\boldsymbol{\delta})]-1/n_2 \sum_i \varphi[\mathbf{Y}_{2j}^*],$$

where $\varphi$ is any non-degenerate measurable non-decreasing function of the data and so $T^*$ corresponds to the comparison of sampling $\varphi$-means: $T^*=\overline{\varphi}_1^* - \overline{\varphi}_2^*$ say. Moreover, they also include non-associative statistics like $T^*(\boldsymbol{\delta})=\varphi[\tilde{\mathbf{Y}}_1^*(\boldsymbol{\delta})]-\varphi[\tilde{\mathbf{Y}}_2^*]$ as for instance the comparison of sampling medians: $T^*=\tilde{\varphi}_1^* - \tilde{\varphi}_2^*$, etc.

Suppose also that effects diverge to the infinity according to whatever monotonic sequence $\{\delta_p, p \geq 1\}$, the elements of which are such that $\delta_p \leq \delta_{p'}$ for any pair $p<p'$. If those conditions are satisfied, then the permutation (conditional) rejection rate of $T$ converges to 1 for all $\alpha$-values not smaller than the minimum attainable $\alpha_a$; thus, $T$ is conditional and unconditional finite-sample consistent. Furthermore, suppose that effects $\boldsymbol{\delta}$ are such that there exists a function $\rho(\boldsymbol{\delta})>0$ of effects $\boldsymbol{\delta}$ the limit of which is 0 as $\boldsymbol{\delta}$ goes to the infinity, $T$ is any test statistic as above, and the data set is obtained by considering the transformation $\mathbf{X}(\boldsymbol{\delta})=\rho(\boldsymbol{\delta})\mathbf{Y}(\boldsymbol{\delta})$. If $\lim_{\delta \uparrow \infty} \boldsymbol{\delta}\rho(\boldsymbol{\delta}) = \tilde{\boldsymbol{\delta}} > 0$, then the unconditional rejection rate converges to 1 for all $\alpha$-values not smaller than the minimum attainable $\alpha_a$; and thus $T$ is weak unconditional finite-sample consistent (Pesarin and Salmaso, 2010b). The extension of these results to random effects $\mathbf{\Delta}$, $0 \leq \Pr\{\mathbf{\Delta}=\mathbf{0}\}<1$, is also shown in Pesarin and Salmaso (2010b).

For instance, suppose a problem in which the $p$-dimensional data set is

$$\mathbf{Y}(\boldsymbol{\delta})= (\delta_k+\sigma_k Z_{k1i},\ i=1,\ldots,n_1;\ \sigma_k Z_{k2i},\ i=1,\ldots,n_2;\ k=1,\ldots,p),$$

where $\delta_k$ and $\sigma_k$ are the fixed effect and the scale coefficient of the $k$-th component variable, respectively, the hypotheses are

$$H_0: \left[\mathbf{Y}_1 \stackrel{d}{=} \mathbf{Y}_2\right] = [\boldsymbol{\delta}=\mathbf{0}] \quad \text{against} \quad H_1: \left[\mathbf{Y}_1 \stackrel{d}{>} \mathbf{Y}_2\right] = [\boldsymbol{\delta}>\mathbf{0}].$$

Suppose also that the test statistic is $T_k^*(\boldsymbol{\delta})=1/p\sum_{k\leq p}[\bar{Y}_{1k}^*(\delta_k)-\bar{Y}_{2k}^*(\delta_k)]/S_k$, where $\bar{Y}_{jk}^*(\delta_h)=\sum_i[Y_{ijk}^*(\delta_h)/n_j$ is the permutation mean of $j$-th sample and $S_k$ a permutation invariant statistic indicator for the $k$-th scale coefficient $\sigma_k$, i.e. a function $S[Y_{ijk}(\delta_k),\ i=1,\ldots,n_j,\ j=1,2]$ of pooled data such as for instance $S_k=\mathbb{M}d[|Y_{ijh}-\tilde{Y}_k|,\ i=1,\ldots,n_j,\ j=1,2]$ the median of absolute deviations from the median specific to the $k$-th variable. It can be proved that a sufficient condition for finite-sample consistency of $T_k^*(\boldsymbol{\delta})$ is that all population means $\mu_k$ exist finite. Thus, when some of the multivariate components do not possess finite mean value, a test based on comparisons of sampling means is not finite-sample consistent. It is worth noting that $T_k^*(\boldsymbol{\delta})$ represents the direct combination (Pesarin and Salmaso, 2010a) of $p$ partial tests $[\bar{Y}_{1k}^*(\delta_k)-\bar{Y}_{2k}^*]$.

In order to extend finite-sample consistency to non-associative statistics, let us briefly introduce the notion of conditional (permutation) unbiasedness for any kind of statistics $T^*(\boldsymbol{\delta})=S(\mathbf{Y}_1^*(\boldsymbol{\delta}))-S(\mathbf{Y}_2^*)$. To this end and with clear meaning of the symbols, let us observe that:

- $T^o(\mathbf{0})=S(\mathbf{Z}_1)-S(\mathbf{Z}_2)$, i.e. the null observed value of statistic $T$.

- $T^o(\pmb{\delta})=S(\mathbf{Z}_1+\pmb{\delta})-S(\mathbf{Z}_2)=S(\mathbf{Z}_1)+D_S(\mathbf{Z}_1,\pmb{\delta})-S(\mathbf{Z}_2)=T^o(0)+D_S(\mathbf{Z}_1,\pmb{\delta})$, where $D_S(\mathbf{Z}_1,\pmb{\delta})\geq 0$.

- $T^*(0)=S(\mathbf{Z}_1^*)-S(\mathbf{Z}_2^*)$, i.e. the value of $T$ in the permutation $\mathbf{u}^*=u_1^*,\ldots,u_n^*$.

- $T^*(\pmb{\delta})=S(\mathbf{Z}_1^*+\pmb{\delta}^*)-S(\mathbf{Z}_2^*)=T^*(0)+D_S(\mathbf{Z}_1^*,\pmb{\delta}^*)-D_S(\mathbf{Z}_2^*)$.

- $D_S(\mathbf{Z}_1^*,\pmb{\delta}^*)\geq D_S(\mathbf{Z}_2^*,0)=0=D_S(\mathbf{Z}_2,0)$, because effects $\delta_{2i}^*$ coming from first group are non-negative.

- $D_S(\mathbf{Z}_1^*,\pmb{\delta}^*)\leq D_S(\mathbf{Z}_1^*,\pmb{\delta})$ point-wise, because in $D_S(\mathbf{Z}_1^*,\pmb{\delta}^*)$ there are non-negative effects assigned to units coming from group 2; e.g., suppose $n_1=3$, $n_2=3$, and $\mathbf{u}^*=(3,5,4,1,2,6)$, then $(\mathbf{Z}_1^*,\pmb{\delta}^*)=[(Z_{13},\delta_{13}),(Z_{22},0),(Z_{21},0)]$, and so

    $$(\mathbf{Z}_1^*,\pmb{\delta})=[(Z_{13},\delta_{13}),(Z_{22},\delta_{11}),(Z_{21},\delta_{12})],$$

    or

    $$(\mathbf{Z}_1^*,\pmb{\delta}_1)=[(Z_{13},\delta_{13}),(Z_{22},\delta_{12}),(Z_{21},\delta_{11})];$$

    it is to be emphasized that $Y(u_i^*)=Z(u_i^*)+\pmb{\delta}(u_i^*)$ if $u_i^*\leq n_1$, that is units coming from first group maintain their effects, whereas the rest of effects are randomly assigned to units coming from second group.

- $D_S(\mathbf{Z}_1^*,\pmb{\delta})\stackrel{d}{=}D_S(\mathbf{Z}_1,\pmb{\delta})$, because $\Pr\{\mathbf{Z}_1^*|\mathcal{X}_{/\mathbf{Y}(0)}\}=\Pr\{\mathbf{Z}_1|\mathcal{X}_{/\mathbf{Y}(0)}\}$ (see Pesarin and Salmaso, 2010a).

Thus $D_S(\mathbf{Z}_1^*,\pmb{\delta}^*)-D_S(\mathbf{Z}_2^*)\leq D_S(\mathbf{Z}_1,\pmb{\delta})$ in permutation distribution and so

$$\begin{aligned}\lambda_T(\mathbf{X}(\pmb{\delta})) &= \Pr\{T(\mathbf{X}^*(\pmb{\delta}))\geq T(\mathbf{X}(\pmb{\delta}))|\mathcal{X}_{/\mathbf{Y}(\pmb{\delta})}\}\\ &= \Pr\{T^*(0)+D_S(\mathbf{Z}_1^*,\pmb{\delta}^*)-D_S(\mathbf{Z}_2^*)-D_S(\mathbf{Z}_1,\pmb{\delta})\geq T^o(0)|\mathcal{X}_{/\mathbf{Y}(0)}\}\\ &\leq \Pr\{T^*(0)\geq T^o(0)|\mathcal{X}_{/\mathbf{Y}(0)}\}=\lambda_T(\mathbf{Y}(0)),\end{aligned}$$

which establishes the dominance in permutation distribution of $\lambda_T(\mathbf{Y}(\boldsymbol{\delta}))$ with respect to $\lambda_T(\mathbf{Y}(0))$, uniformly for all data sets $\mathbf{Y} \in \mathcal{Y}^n$, for all underlying distributions $P$, and for all associative and non-associative statistics $T = S(\mathbf{Y}_1) - S(\mathbf{Y}_2)$.

These results allows us to prove the following:

**Theorem**. *Suppose that in a two-sample problem there are $p \geq 1$ non homoscedastic variables $Y=(Y_1,\ldots,Y_p)$, the observed data set is $\mathbf{Y}(\boldsymbol{\delta}) = (\delta_k + \sigma_k Z_{i1k},\ i=1,\ldots,n_1,\ \sigma_k Z_{i2k},\ i=1,\ldots,n_2;\ k=1,\ldots,p)$ and the hypotheses are*

$$H_0 : \left[ \mathbf{Y}_1 \stackrel{d}{=} \mathbf{Y}_2 \right] = [\boldsymbol{\delta} = \mathbf{0}] \text{ against } H_1 : \left[ \mathbf{Y}_1 \stackrel{d}{>} \mathbf{Y}_2 \right] = [\boldsymbol{\delta} > \mathbf{0}],$$

*where $\boldsymbol{\delta} = (\delta_1, \ldots, \delta_p)'$. For the testing purpose consider the statistic*

$$T_k^*(\boldsymbol{\delta}) = 1/p \sum_{k \leq p} [\tilde{Y}_{1k}^*(\delta_k) - \tilde{Y}_{2k}^*]/S_k,$$

*where $\tilde{Y}_{ji}^*(\boldsymbol{\delta}) = \mathbb{M}d[Y_{ijk}^*(\boldsymbol{\delta})/S_{Mk},\ k=1,\ldots,p]$, $i=1,\ldots,n_j$, $j=1,2$, is the median vector of $p$ scale-free variables specific to $i$-th subject, and $S_{Mk} = MAD_k = \mathbb{M}d[|Y_{ijk} - \tilde{Y}_k|,\ i=1,\ldots,n_j, j=1,2]$ is the median of absolute deviations from the median specific to the variable $Y_k$.*

*In this setting, the test based on $T_{Md}^*(\boldsymbol{\delta})$ is conditional and unconditional finite-sample consistent as far as $p$ diverges and $\mathbb{M}d(Y_1(\boldsymbol{\delta})) > 0$ without requiring existence of any positive moment for $p$ variables.*

**Proof.** For the non-associative statistics it applies the uniformly stochastic ordering of the significance level functions with respect to $\boldsymbol{\delta}$ and $\mathbf{Y}$, that is for $\boldsymbol{\delta}' > \boldsymbol{\delta}$

$$\Pr\{\lambda_T[\mathbf{Y}(\boldsymbol{\delta}')] \leq \alpha\} \stackrel{d}{\leq} \Pr\{\lambda_T[\mathbf{Y}(\boldsymbol{\delta})] \leq \alpha\},$$

hence, with reference to the finite-sample consistency of the second order combined test using the medians

$$T''^{obs} = 1/p\sum_{k\leq p}[\tilde{Y}_{1k}(\delta_k)-\tilde{Y}_{2k}(0)]/S_k = 1/p\sum_{k\leq p}[\tilde{Y}_{1k}(0)+\delta_k-\tilde{Y}^*_{2k}(0)]/S_k =$$

$$= 1/p\sum_{k\leq p}[\tilde{Y}_{1k}(0)-\tilde{Y}_{2k}(0)]/S_k+1/p\sum_{k\leq p}\delta_k/S_k.$$

It should be noted that the quantity $1/p\sum_{k\leq p}[\tilde{Y}_{1k}(0)-\tilde{Y}_{2k}(0)]/S_k$ is nothing else than the arithmetic mean of $p$ sample differences which are all measurable, given that all $p$ involved variables are non-degenerate by assumption (i.e. $S_k>0$; $k=1,...,p$) and, provided that $\min(n_1,n_2)$ is not too small, are all finite (for instance, with the *Pareto* distribution if its parameter is $\gamma \geq [\min(n_1,n_2)/2]$; where [·] is the integer part of (·), the first moment $E_Y(Y,\gamma)$ is finite; it is noticeable that $E_Y(Y,\gamma)$ does not exist $\gamma \leq 1$). Thus, by the law of large numbers for sequences of dependent variables, as $p$ diverges it converges weakly to a constant, not necessarily null. The induced standardized global noncentrality $1/p\sum_{k\leq p}\delta_k/S_k$, which is itself a mean of non-negative and measurable quantities, if it converges, it does so to a positive quantity but it could be let free to diverge as well.

# The permutation approach for ranking of multivariate populations

We formalized our approach to solve the problem we called *the multivariate ranking problem*, i.e. that of ranking several multivariate populations from the 'best' to the 'worst' according to a given pre-specified criterion when a pseudo sample from each population is available and for each marginal univariate response there is a natural preferable direction. Since the key element of our solution is a pseudo testing procedure suitable for multivariate one-sided alternatives, the NPC methodology represents our main methodological reference framework. In fact, to the best of our knowledge, the nonparametric combination of dependent permutation tests, the so-called NPC Tests, is the only method proposed by the literature suitable to achieve this goal. Moreover, when deriving the multivariate one-sided *p*-value-like statistics we can also benefit from the flexibility of the method for obtaining a series of advantages: NPC methodology allows to handle with all type of response variable, i.e. numeric, binary and ordered categorical even in the presence of missing data (at random or not at random, i.e. non-informative or informative) and this can be done also when the number of response variables are much more larger than that of units without the need of having to worry about the curse of dimensionality or the problem of the reduction of degrees of freedom. On the contrary, thanks to the so-called finite-sample consistency of combined permutation tests, the power function does not decrease for any added variable which makes larger standardized noncentrality. It is worth noting that in this situation, which can be common in many real applications, all traditional parametric and nonparametric testing procedures are not at all appropriate (also in the case all multivariate alternatives were of two-sided type). Finally, the NPC approach has a lot of nice feature: it is very low demanding in terms of assumptions and provides always an exact solution for whatever finite sample size whenever the permutation principle applies, i.e. when the pseudo-null hypothesis implies data exchangeability.

We recall that our goal is to classify and ranking *C* multivariate populations with respect to several marginal variables where pseudo samples from each population are available. Note that the multivariate ranking problem is essentially related to a post-hoc comparative multivariate *C*-sample problem where the populations of interest are treatments or groups or items to be investigated by an pseudo experimental or observation study. As we will see later on, although our main reference design will be obviously the one-way MANOVA layout, thanks to the flexibility of the NPC methodology more complex design and analysis are also allowed.

This section is mainly devoted to describe the permutation approach for solving the multivariate ranking problem, from the initial set-up phases to the computation of the final global ranking. In the last part of this section, several simulation studies are presented in order to numerically validate the proposed approach.

## 1.10 Set-up of the multivariate ranking problem

In this section we illustrate in details the steps that must be followed to set up and solve a multivariate ranking problem: first of all we present the different type of designs which may be considered in this context, then in case the ranking process should take account of either confounding factors and/or of intermediate levels of aggregation of the response variables, the so-called stratified and domain analyses are very useful procedures to be applied in these situations. Finally, we point up on some more practical questions: the choices of the test statistics and of the combining function and the specific pairwise permutation strategy to be used. We close with some issues on the multiplicity control and simultaneous testing.

### 1.10.1 Types of designs

Up to now the main reference design for the multivariate ranking problem has been the so-called one-way MANOVA layout whose statistical model in case of fixed effects can be represented as:

$$\mathbf{Y}_{ij} = \boldsymbol{\mu}_j + \boldsymbol{\varepsilon}_{ij} = \boldsymbol{\mu} + \boldsymbol{\tau}_j + \boldsymbol{\varepsilon}_{ij}, \; i=1,\ldots,n_j, j=1,2,\ldots,C, \tag{9}$$

where $\boldsymbol{\mu}_j$ (or $\boldsymbol{\tau}_j$) is the *p*-dimensional mean effect, $\boldsymbol{\varepsilon}_{ij} \sim \text{IID}(0,\boldsymbol{\Sigma})$ is a *p*-variate pseudo random term of experimental errors with zero mean and variance/covariance matrix $\boldsymbol{\Sigma}$.

Anyway extensions to more complex designs and situations are also possible. In fact, either the Multivariate Randomized Complete Block – MRCB design, the Repeated Measures – RM design and finally the multivariate *C*-sample comparison of curves or trajectories (functional data), can be taken into consideration:

- MRCB design: $\mathbf{Y}_{ij} = \boldsymbol{\mu} + \boldsymbol{\tau}_j + \boldsymbol{\beta}_i + \boldsymbol{\varepsilon}_{ij}$, where $\boldsymbol{\beta}_i$, $i=1,...,n_j$, is the block effect (see Aboretti et al., 2012);

- RM design/functional data (with random effects): $\mathbf{Y}_{ji}(t) = \boldsymbol{\mu} + \boldsymbol{\eta}_j(t) + \boldsymbol{\varepsilon}_{ji}(t)$,

  where $\boldsymbol{\varepsilon}_{ji}(t) = \boldsymbol{\Delta}_{ji}(t) + \boldsymbol{\sigma}(\boldsymbol{\eta}_j(t)) \cdot \mathbf{Z}_{ji}(t)$, $i=1,...,n_j$, $j=1,...,C$, $t=1,...,N$, where $\mathbf{Z}_{ji}(t)$ are non-Gaussian error terms distributed as a stationary stochastic process with null mean and unknown distribution $P_\mathbf{Z}$, $\boldsymbol{\mu}$ is a population constant, coefficients $\boldsymbol{\eta}_j(t)$ represent the *main treatment effects* and may depend on time through any kind of function, but are independent of units; quantities $\boldsymbol{\Delta}_{ji}(t)$ represent the so-called *individual effects*, and $\boldsymbol{\sigma}(\boldsymbol{\eta}_j(t))$ are time-varying scale coefficients which may depend, through monotonic functions, on main treatment effects $\boldsymbol{\eta}_j$.

From the point of view of obtaining a valid pseudo testing solution and within the NPC framework, all situations listed above are no more than designs with nuisance parameters ($\boldsymbol{\tau}$, $\boldsymbol{\eta}$ and $\boldsymbol{\Delta}$) that can be fully removed exploiting the concept of constrained permutations (Basso etl all, 2009; Pesarin and Salmaso, 2010): since under the null hypothesis the exchangeability of observations holds only within given conditions, the null permutation distributions of the test statistics are computed allowing permutations to occur only under a given restriction. In practice, once a suitable blocking factor is defined and permutations are allowed only among samples within the same level of that blocking factor when calculating the pseudo test statistics it happens that all nuisance parameter are implicitly removed (they vanish by applying suitable linear transformations, see Basso et. all, 2009).

Note that in case of even more complex designs such as the Latin and Graeco-Latin squares (Montgomery, 2012), the concept just exposed remains valid as long as the two or more blocking factors are used so as to define a single blocking factor whose levels are obtained from the combinations of the levels of the individual blocking factors.

### 1.10.2 Stratified analysis

Sometimes in the multivariate ranking problem we should take the presence of possible confounding factors into consideration that is there are units/subjects features which potentially have a noise effect on the problem at hand. Typical confounding factors are sex or age of the subjects. In this situation a stratified analyses can be very useful in order to provide, before getting a final global ranking, separated results for each level of the stratification factor. In practice, to handle with a stratification factor we use an additional classification criterion of units/subjects and then we allow permutations among samples only within the same level of stratification factor.

Note that even if blocking and stratification are both handled by a restriction on the exchangeability under the null hypothesis, only apparently they appear as the same situation. In fact, blocking refers to a technique aimed at removing nuisance parameters, while stratification provides intermediate separated results for each stratum and a final global analysis where the possible confounding effect has been removed thanks to a constrained permutation strategy.

### 1.10.3 Domain analysis

We refer to a domain as a result of a classification or grouping of marginal response variables which share some basic features with respect to the problem at hand. For examples, in shape analysis domains are subgroups of landmarks sharing anatomical, biological or locational features (Brombin and Salmaso, 2009). Very often in a multivariate problem we are facing the presence of such kind of domains, let us think for example on the sections of questionnaire on the consumer relevance of a product (example 5.4) or on the type of stain of a primary performance experiment in laundry industry (example 5.1). In a similar way as in the stratified analysis, the presence of domains in the multivariate ranking problem suggests to provide intermediate results for each domain before to obtain a final global analysis.

### 1.10.4 Choice of the test statistics

It is worth noting that within the NPC framework an optimal statistic cannot exist because it is function of the population distributions which is unknown by definition (Pesarin and Salmaso, 2010). For this reason it is important to consider for each type of response variable a number of different test statistics. We recall that each univariate partial pseudo test statistic we are presenting must be suitable for one-sided alternatives with respect to the hypotheses $H_{0k(jh)}$ vs. $H_{1k(jh)}$:

$$\begin{cases} H_{0k(jh)}: Y_{jk} \stackrel{d}{=} Y_{hk} \\ H_{1k(jh)}: \left(Y_{jk} \stackrel{d}{<} Y_{hk}\right) \bigcup \left(Y_{jk} \stackrel{d}{>} Y_{hk}\right) \end{cases}, j,h = 1,...,C, j \neq h, k = 1,..,p \ .$$

As far as the multivariate test statistics suitable for testing the hypotheses $H_{0(jh)}$ vs. $H_{k(jh)}$ is concerned,

$$\begin{cases} H_{0(jh)}: \mathbf{Y}_j \stackrel{d}{=} \mathbf{Y}_h \\ H_{1(jh)}: \left(\mathbf{Y}_j \stackrel{d}{<} \mathbf{Y}_h\right) \bigcup \left(\mathbf{Y}_j \stackrel{d}{>} \mathbf{Y}_h\right) \end{cases}, j,h = 1,...,C, j \neq h,$$

we will apply the nonparametric combination methodology using a suitable combining function.

#### 1.10.4.1 **Continuous or binary response variables**

When the univariate marginal response variable is continuous or binary, within the permutation framework we can use a number of test statistics suitable for one-sided alternatives. In this context, we underline that the test statistics should obviously not permutationally equivalent, in particular we can refer to

- difference of sample means;

- difference of sample means in case of missing values;
- difference of sample medians or quartiles.

When providing the above list, we implicitly assumed that the ranking problem was a location-type problem where a natural preference direction for the response variable does exist. Actually, if the ranking criteria would be based on the classification of several populations with respect to the variability (scale ranking problem), we may consider a number of scale-type test statistics:

- difference of sample standard deviations;
- difference of sample interquartile ranges.

### 1.10.4.2     Ordered categorical response variables

When the univariate marginal response variable is ordered categorical, within the permutation framework we can use a number of test statistics suitable for directional alternatives:

- Anderson-Darling;
- Multi-focus;
- Difference of mid-ranks;
- Kolmogorov-Smirnov.

As in case of continuous or binary response, when providing the above list we implicitly assumed that the ranking problem was similar to a shift-in-distribution problem where a natural preference direction for the ordered categorical response variable does exist. Actually, if the ranking criterion would be based on the classification of several populations with respect to the variability (heterogeneity ranking problem), we may consider a number of scale-type test statistics:

- difference of sample Shannon or Gini indexes.

### 1.10.4.3     Multi-aspect

It is worth noting that within the NPC framework an optimal statistic cannot exist because it is function of the population distributions which is unknown by definition (Pesarin and Salmaso, 2010a). More formally, when the whole data set **Y** is minimal sufficient in $H_0$, univariate statistics suitable for summarizing the whole information on an aspect of interest do not exist. To overcome this limitation and in order to reduce the loss of information associated with using only one single overall statistic, it is possible to take account of a set of statistics suitable for complementary or concurrent view-points, each fitted for summarizing information on a specific aspect of interest for the problem, and so to find solutions within the so-called *multi-aspect strategy* (Salmaso and Solari, 2005), i.e. combining different test statistics, suitable for testing different aspects related to the same univariate null hypothesis by working on the same dataset. The multi-aspect strategy was originally proposed by R.A. Fisher: "In hypotheses testing problems the experimenter might have to face not only one, but a class of hypotheses, and it may happen that he is interested in all the hypotheses of such class… It follows that different significance tests may be thought as a set of tests for testing different aspects of the same hypothesis" (Pesarin and Salmaso, 2010). Hence, since different test statistics may be suitable and effective for testing different aspects of the same null hypothesis (i.e. univariate location problem), instead of using just one statistic per variable we may use a list of statistics and then combine all of them to get a final multivariate and multi-aspect *p*-value-like statistic. In this way we obtain two advantages: at first, by using several pseudo test statistics, this allows us to possibly include the more sensitive procedures with respect to the unknown alterative hypothesis and population distributions and then, thanks to the NPC methodology, we have the chance to gain more additional power (Pesarin and Salmaso, 2010b) by using a suitable multi-aspect procedure.

### 1.10.5 Choice of the pairwise permutation strategy

Under the null hypothesis $H_{0k(jh)}$ of equality of the *j*-th and *h*-th populations with respect to the *k*-th response variable, it is worth noting that there are actually four different but all valid strategies which we can take into account when calculating the conditional permutation space of the permutation test statistics $T_{k(jh)}$:

- pairwise independent permutations - PIP;

- not pairwise permutations (C sample design) - NPIP;

- pairwise constrained synchronized permutations - PCSP;

- pairwise unconstrained synchronized permutations - PUSP.

The first two strategies are allowed either in case of balanced and of unbalanced designs, while the latter two strategies are valid only in case of balanced design.

Pairwise independent permutations means that for each ($j,h$)-th pairwise comparison we perform independent permutations, while not pairwise permutations refers to usual permutation strategy usually applied for the *C*-sample permutation test, where permutations are performed among all the *C*-sample data (Pesarin and Salmaso, 2010). In the situations, the cardinality of the permutation test statistic $T_{k(jh)}$ is defined as

- $S(\text{PIP})_{(j,h)} = \binom{n_i + n_h}{n_i}$;

- $S(\text{NPIP})_{(j,h)} = \frac{n!}{n_1!...n_C!}$;

where $j,h =1,2,...,C, j \neq h$. Note that $S$(NPIP) is much more larger than $S$(PIP), in practice within the permutation framework calculating the null distribution of the pairwise pseudo test statistic with reference to the *C*-sample null hypothesis instead of the more natural two-sample null hypothesis allows us to enlarge the support of the permutation pseudo test statistic $T_{k(jh)}$. Obviously there is an underlying drawback: in case the true differences are located in the most (or in all) pairwise comparisons the NPIP strategy may result in a loss of power while it can be very useful when there are only a few pairwise comparisons under the alternative hypothesis.

With the term "synchronized permutations" we mean that the permutation strategy is defined so that permutations are allowed to occur only under a given restriction constraint which is needed to remove some nuisance parameters (such as in the case of blocking or stratification). For example when testing the main effects in the two way ANOVA using permutation tests, permutations are allowed to occur only within levels of the not under testing factor while permutations respect to the under testing factor are not allowed (Basso et al., 2009).

Referring to the case of balanced designs, for both pairwise constrained and unconstrained synchronized permutations, the cardinality of the permutation test statistics $T_{k(jh)}$ is defined as

- $S(\text{PCSP})_{(j,h)} = S(\text{PUPS})_{(j,h)} = \binom{2n}{n} = \frac{2n!}{n!n!}$.

It is interesting to highlight that even in case of balanced designs it happens that $S(\text{PIP})_{(j,h)} = S(\text{PCSP})_{(j,h)} = S(\text{PUPS})_{(j,h)}$, actually the cardinality of the multivariate test statistic $T^+_{(jh)}$ we use as multivariate score to test the alternative

$$H^+_{1(j\bullet)} : \bigcap_h \left( \mathbf{Y}_j \stackrel{d}{>} \mathbf{Y}_h \right), h = 1,...,C, j \neq h,$$

has a different cardinality with respect to each one of the three specific permutation strategies. The reason is due to the fact that from the one hand the PCSP and PUPS constraint all pairwise comparison to have the same (for PCSP) or 'similar' (for PUPS) permutations (for details see Basso et al., 2009), while on the other hand the PIP strategy exploiting the independence of permutations with respect to the individual pairwise comparisons allows us to obtain a larger cardinality of the permutation space of $T^+_{(jh)}$, in particular it can be proved that $S(\text{PIP})_{T^+_{(j,h)}} > S(\text{PUPS})_{T^+_{(j,h)}} > S(\text{PCPS})_{T^+_{(j,h)}}$.

### 1.10.6 Choice of the combining function

In order to define one-sided multivariate pseudo test statistics within the combination of dependent permutation tests methodology, a suitable combining function must be chosen (Pesarin and Salmaso, 2010). Frequently used combining function are:

- Fisher combination: $\phi_F = -2 \sum_k \log(\lambda_k)$;

- Tippet combination: $\phi_T = \max_{1 \leq k \leq p} (1 - \lambda_k)$;

- Direct combination: $\phi_D = \sum_k T_k$;

- Liptak combination: $\phi_L = \sum_k \Phi^{-1}(1-\lambda_k)$;

where $k = 1,\ldots,p$ and $\Phi$ is the standard normal c.d.f.

It can be seen that under the global null hypothesis the CMC procedure allows for a consistent estimation of the permutation distributions, marginal, multivariate and combined, of the $k$ partial tests. Usually, Fisher's combination function is considered (Pesarin and Salmaso, 2010), mainly for its finite and asymptotic good properties. Of course, it would be also possible to take into consideration any other combining function (Lancaster, Mahalanobis, etc.; see Folks, 1984). The combined test is also unbiased and consistent. For a detailed description we refer to Pesarin and Salmaso (2010).

It is worth noting that since within the NPC framework an optimal combination function in general does not exist because each combining function has different sensitivity to different configurations of the alternative hypothesis, that is every combining function has its own characteristics that makes it preferable instead of another in a specific situation. In order to better understand this concept, let us consider the critical regions of the three combining functions in the case of two independent tests (Figure 2).

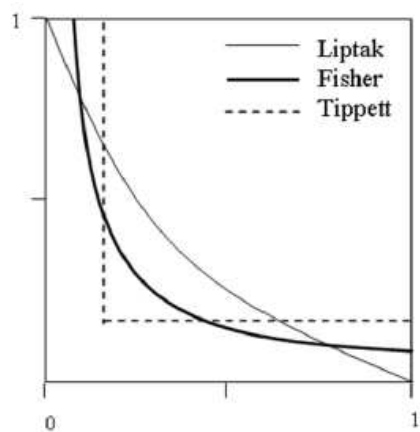

*Figure 2. Critical regions of three combining functions in the case of two independent tests.*

Figure 2 suggests that inferential results may slightly differ using different combining functions. In order to reduce the effect of this limitation from the computational point of view, we can try to iteratively apply different combination functions in order to reach a more stable result rather than applying one specific combining function. This is the so-called *iterated combination strategy*, formally described by the following algorithm:

1. apply at least three different combining functions (e.g. Fisher, Liptak and Tippett) to the same partial *p*-values; in general, the obtained *p*-values will be slightly different;

2. apply to results of step 1 the same combining functions; in general, the obtained *p*-values will be different but slightly closer one another;

3. iteratively repeat step 2 until all combining functions provide slightly the same resulting *p*-value.

Figure 3 reports an example of the behaviour of the iterated combination, showing that in this specific case after six iterations the resulting *p*-value is practically independent of the choice of the specific combining function.

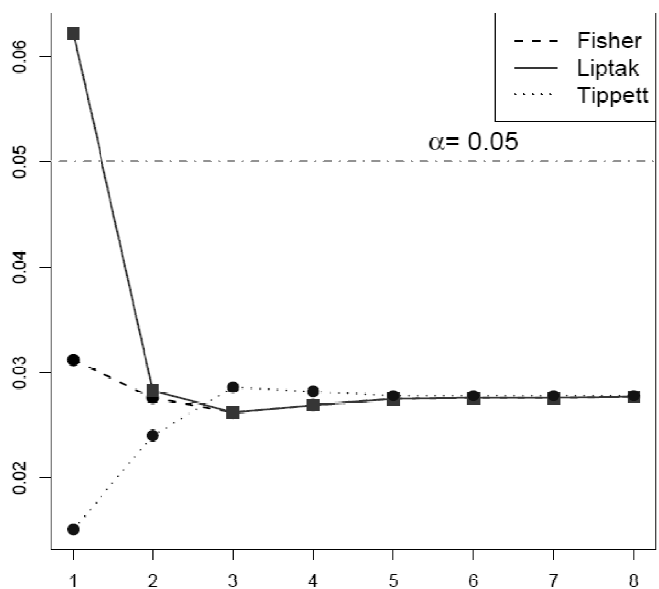

*Figure 3. Behavior of the Iterated Combination.*

### 1.10.7 Multiplicity issue and simultaneous testing

The multiplicity issue occurs in case of simultaneous testing when a set, or family, of statistical tests is considered simultaneously. Since the multivariate ranking problem make use of pseudo inferential tools, obviously if falls within this context either because we are facing a multivariate response and we are considering a set of multiple comparisons procedure (MCPs). However, within the multivariate ranking problem the need to properly control of the global type I error actually occurs only at the final stage, that is when a set of $C \times (C-1)$ one-sided pairwise comparisons are performed via *p*-value-like statistics.

In general, incorrect rejection of the null hypothesis is more likely when the family as a whole is considered and failure to compensate for multiple comparisons can can lead to committing serious mistakes in the process of classification and ranking. Among the many definitions of global type-I error for MCPs, we take into consideration the most used, i.e. Familywise Error Rate (FWER), which is the probability of rejecting at least one true null sub-hypothesis (for a review of alternative definitions and their main properties see Westfall et al., 2011).

Several statistical techniques have been developed to prevent this problem, mainly adjusting the significance levels of the considered tests. The so-called single step procedures work singularly on each sub-test and do not take the dependence structure of the tests into account. Among the most famous and most used methods are Bonferroni and Tukey solutions. The stepwise procedures firstly test only some sub-hypotheses and according to the results, they then take other subhypotheses into consideration, until a given condition is satisfied. Among the most used stepwise procedures are Bonferroni-Holm method and the Shaffer proposals (1986).

### 1.11 From one-sided multivariate pairwise comparisons to global ranking

Once the set of $C \times (C-1)$ multivariate one-sided pseudo permutation *p*-values $P_\bullet^+$ are computed, then rows and columns of the matrix $P_\bullet^+$ are re-arranged into the matrix $P_{[\bullet]}^+$ according to a suitable statistic $p^+_{\bullet(j\bullet)}$, $j=1,...,C$, monotonically related with multivariate distances (for details see section 1.2 Formalization of the problem and general solution), the next step is to use this statistic to estimate the global ranking of the *C* multivariate populations under investigation, i.e. classify the populations from the 'best' to the 'worst'. We refer to this ranking as 'global ranking' to underline the fact it is devoted to rank these populations from a multivariate point of view. From simple inspection of the re-arranged one-sided pseudo *p*-values $P^+_{[\bullet]}$, there are a few situations in which the definition of a global ranking can be easily and unequivocally established:

− when all pairwise differences are significant, that is all $p^+_{\bullet[jh]}$, $j,h=1,...,C, j\neq h$, are lower than the desired significance $\alpha$–level; for example, setting $\alpha=5\%$, Table 6 highlights two situations in case of three and four populations respectively, where the final global ranking always matches and confirms the preliminary classification of populations [1],[2],...[*C*];

− when the significant differences appear in blocks, i.e. there are sort of 'clusters' of populations and within the cluster the populations will obviously have the same rank (see Table 7, where blocks are highlighted with diagonal lines).

*Table 6. Two examples of all significant pairwise differences in case of C=3 and 4.*

| $P^+_{[\bullet]}=$ | | [P1] | [P2] | [P3] |
|---|---|---|---|---|
| | [P1] | - | .0009 | .0002 |
| | [P2] | - | - | .0003 |
| | [P3] | - | - | - |
| | ranking: | 1 | 2 | 3 |

| $P^+_{[\bullet]}=$ | | [P1] | [P2] | [P3] | [P4] |
|---|---|---|---|---|---|
| | [P1] | - | .0005 | .0007 | .0007 |
| | [P2] | - | - | .0006 | .0003 |
| | [P3] | - | - | - | .0003 |
| | [P4] | - | - | - | - |
| | ranking: | 1 | 2 | 3 | 4 |

*Table 7. Several examples of significant pairwise differences in blocks in case of C=3 and 4.*

| $P^+_{[\bullet]}=$ | | [P1] | [P2] | [P3] |
|---|---|---|---|---|
| | [P1] | - | .4569 | .0002 |
| | [P2] | - | - | .0003 |
| | [P3] | - | - | - |
| | ranking: | 1 | 1 | 3 |

| $P^+_{[\bullet]}=$ | | [P1] | [P2] | [P3] | [P4] |
|---|---|---|---|---|---|
| | [P1] | - | .7225 | .0017 | .0006 |
| | [P2] | - | - | .0016 | .0003 |
| | [P3] | - | - | - | .3239 |
| | [P4] | - | - | - | - |
| | ranking: | 1 | 1 | 3 | 3 |

| $P^+_{[\bullet]}=$ | | [P1] | [P2] | [P3] | [P4] |
|---|---|---|---|---|---|
| | [P1] | - | .0035 | .0027 | .0007 |
| | [P2] | - | - | .6269 | .5718 |
| | [P3] | - | - | - | .1527 |
| | [P4] | - | - | - | - |
| | ranking: | 1 | 2 | 2 | 2 |

66As we can easily guess by the above tables, in these situations we can directly derive the global ranking from a simple visual inspection of the pseudo *p*-values of matrix $P_{[\bullet]}^{+}$.

### 1.11.1 The issue of logical relations among hypotheses being tested

However, since in general the above listed situations, where the global ranking is easy and straightaway to derive, are frequently not met, we need to think on a meaningful criterion to properly rank populations using the pseudo *p*-value results. In fact, it is well known that within the multiple comparisons procedures a sort of transitive property of significant differences obviously does not apply, so that when interpreting the pairwise results very often it happens that we encounter apparent logical incoherencies as pointed out by Shaffer (1986). In other words, we are referring to the possible conflict between the results that emerge from the multiple comparisons and the logic behind the hypotheses that are tested. As a matter of fact, the hypotheses being tested are always logically interrelated so actually not all combinations of true and false hypotheses are possible. As a simple example, in case of three populations $P_1$, $P_2$ and $P_3$ it is easily seen from the relations among hypothesis that if $H_{0(jh)}$: $P_j = P_h$ is false for any given pair $j \neq h$, $j,h=1,...,C$, at least one other remaining hypotheses must be false. Thus, in case of three populations *from the logical point of view* there cannot be one false and two true hypotheses among these three but *from the pairwise results point of view* it can happen that only one *p*-value is lower than the significance $\alpha$–level, that is we can observe just one rejection and two acceptances. In case of three and four populations, Table 8 highlights some examples of this kind of conflicting decisions related to the global ranking definition where possible incoherent pseudo *p*-values are highlighted in grey (while with diagonal lines are denoted the blocks that conflict with the grey cells).

*Table 8. Several examples of possible conflicting decision on ranking in case of C=3 and 4.*

| $P_{[\bullet]}^{+}=$ | | P1 | P2 | P3 | | $P_{[\bullet]}^{+}=$ | | P1 | P2 | P3 | P4 | | $P_{[\bullet]}^{+}=$ | | P1 | P2 | P3 | P4 |
|---|---|---|---|---|---|---|---|---|---|---|---|---|---|---|---|---|---|---|
| | P1 | - | .4889 | .0442 | | | P1 | - | .6057 | .3971 | .0007 | | | P1 | - | .2532 | .0371 | .0279 |
| | P2 | - | - | .3410 | | | P2 | - | - | .0326 | .0243 | | | P2 | - | - | .1639 | .0333 |

| P3 | - | - | - |
|----|---|---|---|

| P3 | - | - | - | 0394 |
|----|---|---|---|------|
| P4 | - | - | - |      |

| P3 | - | - | - | 0909 |
|----|---|---|---|------|
| P4 | - | - | - |      |

## 1.11.2 The ranking algorithm

To overcome the just mentioned problem of possible apparent logical incoherencies that can arise from significances of pseudo multivariate one-sided pairwise *p*-values when trying to derive from these results the classification and ordering of populations into a final global ranking, we propose a solution which is inspired by the so-called *underlining* which is one of the a graphical representations of Tukey's (1953) and Tukey-Kramer's (1956) methods for multiple comparisons (Hsu and Peruggia, 1994). The Tukey's *underlining* prescribes that after ordering the populations according to the increasing values of their estimated means, all subgroups of populations that cannot be declared different are underlined by a common line segment.

Our proposed algorithm from the one hand is designed to formalize the logic underlying the Tukey's *underlining* procedure and from the other hand it aims to extend that idea from the multivariate point of view. The algorithm consists of the following steps:

1. order from the lowest to the highest the *C* one-sided pseudo *p*-values $p^+_{\bullet(j\bullet)}$, *j*=1,...,*C*, into $p^+_{\bullet[j\bullet]}$; we recall that $p^+_{\bullet(j\bullet)}$ represents a pseudo-inferential statistic monotonically related with multivariate distances among populations; this process provides a preliminary classification of populations [1],[2],...[*C*]: at the top there will be the tentative 'best' population, at the second place the tentative 'best' population among the remaining *C*−1 and so on;

2. apply the same ranking [1],....[C] to the rows and columns of the matrix $P^+_\bullet$ containing the (unordered) multivariate directional pseudo *p*-values $p^+_{\bullet(jh)}$, this means that we re-arrange the rows and columns of the matrix $P^+_\bullet$ to obtain the matrix $P^+_{[\bullet]}$:

$$P_{\bullet}^{+} = \begin{bmatrix} - & p_{\bullet(12)}^{+} & p_{\bullet(13)}^{+} & \cdots & p_{\bullet(1C)}^{+} \\ p_{\bullet(21)}^{+} & - & p_{\bullet(23)}^{+} & \cdots & p_{\bullet(2C)}^{+} \\ \cdots & \cdots & - & \cdots & \cdots \\ p_{\bullet(C-1,1)}^{+} & p_{\bullet(C-1,2)}^{+} & \cdots & - & p_{\bullet(C-1,C)}^{+} \\ p_{\bullet(C1)}^{+} & p_{\bullet(C2)}^{+} & \cdots & p_{\bullet(C,C-1)}^{+} & - \end{bmatrix} \Rightarrow P_{[\bullet]}^{+} = \begin{bmatrix} - & p_{\bullet[12]}^{+} & p_{\bullet[13]}^{+} & \cdots & p_{\bullet[1C]}^{+} \\ p_{\bullet[21]}^{+} & - & p_{\bullet[23]}^{+} & \cdots & p_{\bullet[2C]}^{+} \\ \cdots & \cdots & - & \cdots & \cdots \\ p_{\bullet[C-1,1]}^{+} & p_{\bullet[C-1,2]}^{+} & \cdots & - & p_{\bullet[C-1,C]}^{+} \\ p_{\bullet[C1]}^{+} & p_{\bullet[C2]}^{+} & \cdots & p_{\bullet[C,C-1]}^{+} & - \end{bmatrix};$$

3. remove the lower-diagonal elements of $P_{[\bullet]}^{+}$ to define a final upper diagonal pseudo *p*-value matrix $P_{[\bullet]upper}^{+}$:

$$P_{[\bullet]}^{+} = \begin{bmatrix} - & p_{\bullet[12]}^{+} & p_{\bullet[13]}^{+} & \cdots & p_{\bullet[1C]}^{+} \\ p_{\bullet[21]}^{+} & - & p_{\bullet[23]}^{+} & \cdots & p_{\bullet[2C]}^{+} \\ \cdots & \cdots & - & \cdots & \cdots \\ p_{\bullet[C-1,1]}^{+} & p_{\bullet[C-1,2]}^{+} & \cdots & - & p_{\bullet[C-1,C]}^{+} \\ p_{\bullet[C1]}^{+} & p_{\bullet[C2]}^{+} & \cdots & p_{\bullet[C,C-1]}^{+} & - \end{bmatrix} \Rightarrow P_{[\bullet]upper}^{+} = \begin{bmatrix} - & p_{\bullet[12]}^{+} & p_{\bullet[13]}^{+} & \cdots & p_{\bullet[1C]}^{+} \\ - & - & p_{\bullet[23]}^{+} & \cdots & p_{\bullet[2C]}^{+} \\ - & - & - & \cdots & \cdots \\ - & - & - & - & p_{\bullet[C-1,C]}^{+} \\ - & - & - & - & - \end{bmatrix};$$

4. using for example the method proposed by Shaffer (1986), adjust by multiplicity the set of the one-sided pseudo *p*-values $p_{\bullet[jh]}^{+}$, i.e. the elements of $P_{[\bullet]upper}^{+}$, into $P_{[\bullet]adj}^{+}$;

5. referring to a desired significance $\alpha$–level, transform the adjusted pseudo *p*-values $p_{\bullet[jh]adj}^{+}$ into 0-and-1 values where each element $s_{[jh]}$ takes the value of 1 if $p_{\bullet[jh]adj}^{+} > \alpha$, otherwise it takes 0 if $p_{\bullet[jh]adj}^{+} \leq \alpha$;

6. multiply each element $s_{[jh]}$ of *S* by the value *j*, that is the *j*-th row of *S* where the element $s_{[jh]}$ is lying; note that implicitly *j* represents the preliminary classification of populations performed in step 1; in this way we define a rank score $r_{[jh]} = s_{[jh]} \times j$ whose elements are put in the matrix *R*;

7. calculate along the columns the average of rank scores $r_{[jh]}$: $\bar{r}_{[j]} = \dfrac{\sum_{h=1}^{j} r_{[jh]}}{j}$ obtaining in this way a final ranking global score $\bar{r}_{[j]}$; finally, by applying the rank transformation on this scores we obtain the global ranking.

The next scheme summarizes the last three steps of the proposed algorithm.

$$S = \begin{bmatrix} - & s_{12} & s_{13} & \cdots & s_{1C} \\ - & - & s_{23} & \cdots & s_{2C} \\ - & - & - & \cdots & \cdots \\ - & - & - & - & s_{C-1,C} \\ - & - & - & - & - \end{bmatrix} \Rightarrow$$

$$R = \begin{bmatrix} - & s_{12} & s_{13} & \cdots & s_{1C} \\ - & - & s_{23} & \cdots & s_{2C} \\ - & - & - & \cdots & \cdots \\ - & - & - & - & s_{C-1,C} \\ - & - & - & - & - \end{bmatrix} \otimes \begin{bmatrix} 1 \\ 2 \\ \cdots \\ C \\ - \end{bmatrix} = \begin{bmatrix} - & r_{12} & r_{13} & \cdots & r_{1C} \\ - & - & r_{23} & \cdots & r_{2C} \\ - & - & - & \cdots & \cdots \\ - & - & - & - & r_{C-1,C} \\ - & - & - & - & - \\ \overline{r_1} & \overline{r_2} & \cdots & & \overline{r_C} \end{bmatrix}.$$

In order to better understand this procedure, we report in Figure 4 an example of application of the above algorithm with $C = 8$ populations.

| Ord. tr.  | (1) | (2) | (3) | (4) | (5) | (6) | (7) | (8) |
|-----------|-----|-----|-----|-----|-----|-----|-----|-----|
| (1)       | 1   | 1   | 1   | 0   | 0   | 0   | 0   | 0   |
| (2)       |     | 1   | 1   | 1   | 1   | 0   | 0   | 0   |
| (3)       |     |     | 1   | 1   | 1   | 1   | 0   | 0   |
| (4)       |     |     |     | 1   | 1   | 1   | 1   | 0   |
| (5)       |     |     |     |     | 1   | 1   | 1   | 1   |
| (6)       |     |     |     |     |     | 1   | 1   | 1   |
| (7)       |     |     |     |     |     |     | 1   | 1   |
| (8)       |     |     |     |     |     |     |     | 1   |
| Rank ass. |     |     |     |     |     |     |     |     |
| 1         | 1   | 1   | 1   |     |     |     |     |     |
| 2         |     | 2   | 2   | 2   | 2   |     |     |     |
| 3         |     |     | 3   | 3   | 3   | 3   |     |     |
| 4         |     |     |     | 4   | 4   | 4   | 4   |     |
| 5         |     |     |     |     | 5   | 5   | 5   | 5   |
| Col.s mean| 1   | 1.5 | 2   | 3   | 3.5 | 4   | 4.5 | 5   |
| Rank      | 1   | 2   | 3   | 4   | 5   | 6   | 7   | 8   |

*Figure 4. Example of application of the global ranking algorithm with* $C = 8$ *treatments.*

The rationale behind this algorithm is very simple and intuitive: after pre-ordering the populations using a suitable statistic monotonically related with multivariate distances, we simply estimate the rank of each population by the average of all possible ranks assigned to that population. Note that if all pairwise differences were significant or when significant differences would appear in blocks, the algorithm will exactly provide the same global ranking which could be directly be derived by visual inspection of the pseudo *p*-values of matrix $P_{[\bullet]}^+$ (see Table 6 and Table 7).

As it is evident from the steps of our algorithm, the analogies with the Tukey's *underlining* method (1953) are as follows:

- at first, with the same Tukey's logic behind the ordering of univariate populations using the sample means, we also order the population but according to a *multivariate score* (because our problem is multivariate in nature);

- the transformation of pseudo multivariate significance directional differences into a 1-and-0 matrix is nothing more than a formal representation of the process of Tukey's underlining, of which we do a final summary by calculating the score average along the matrix columns. In other words, we 'underline' a subgroup of not significant populations by assigning them the same rank.